\documentclass[twocolumn]{aastex631}

\usepackage{xcolor}

\newcommand{\lya}{Ly$\alpha$}
\newcommand{\fesclya}{$f_{esc}^{Ly\alpha}$}
\newcommand{\fesc}{$f_{esc}^{LyC}$}
\newcommand{\ha}{H$\alpha$}

\newcommand{\hi}{H\textsc{i}}

\begin{document}

\title{The Ly$\alpha$ and Continuum Origins Survey. III. Investigating the link between galaxy morphology, merger properties and LyC escape}

\author[0000-0003-1767-6421]{Alexandra Le Reste}
\affiliation{Minnesota Institute for Astrophysics, University of Minnesota, 116 Church Street SE, Minneapolis, MN 55455, USA}

\author{Anne E. Jaskot}
\affiliation{Department of Astronomy, Williams College, Williamstown, MA 01267, USA}

\author{Jordanne Brazie}
\affiliation{Department of Astronomy, Williams College, Williamstown, MA 01267, USA}

\author[0000-0002-9136-8876]{Claudia Scarlata}
\affiliation{Minnesota Institute for Astrophysics, University of Minnesota, 116 Church Street SE, Minneapolis, MN 55455, USA}

\author[0000-0002-0159-2613]{Sophia R. Flury}
\affiliation{Institute for Astronomy, University of Edinburgh, Royal Observatory, Edinburgh, EH9 3HJ, UK}

\author[0000-0002-6016-300X]{Kameswara B. Mantha}
\affiliation{Minnesota Institute for Astrophysics, University of Minnesota, 116 Church Street SE, Minneapolis, MN 55455, USA}
\affiliation{Missouri Institute of Technology, University of Missouri-Kansas City, 5110 Rockhill road, Kansas City, MO, 64110.}

\author{Alaina Henry}
\affiliation{Center for Astrophysical Sciences, Department of Physics \& Astronomy, Johns Hopkins University, Baltimore, MD 21218, USA}
\affiliation{Space Telescope Science Institute; 3700 San Martin Drive, Baltimore, MD 21218, USA}

\author[0000-0001-8587-218X]{Matthew J. Hayes}
\affiliation{Department of Astronomy, Oskar Klein Centre, Stockholm University,106 91 Stockholm, Sweden}

\author[0000-0002-3005-1349]{G\"{o}ran \"{O}stlin}
\affiliation{Department of Astronomy, Oskar Klein Centre, Stockholm University,106 91 Stockholm, Sweden}

\author[0000-0001-8419-3062]{Alberto Saldana-Lopez}
\affiliation{Department of Astronomy, Oskar Klein Centre, Stockholm University,106 91 Stockholm, Sweden}

\author[0000-0001-5331-2030]{Trinh X. Thuan}
\affiliation{Astronomy Department, University of Virginia, P.O. Box 400325, Charlottesville, VA 22904-4325,USA}

\author[0000-0002-6849-5375]{Maxime Trebitsch}
\affiliation{LUX, Observatoire de Paris, Université PSL, Sorbonne Université, CNRS, 75014 Paris, France}

\author[0000-0002-9217-7051]{Xinfeng Xu}
\affiliation{Department of Physics and Astronomy, Northwestern University,2145 Sheridan Road, Evanston, IL, 60208, USA.}
\affiliation{Center for Interdisciplinary Exploration and Research in Astrophysics (CIERA), 1800 Sherman Avenue, Evanston, IL, 60201, USA.}

\author[0000-0001-5758-1000]{Ricardo O. Amor\'{i}n}
\affiliation{Instituto de Astrof\'{i}sica de Andaluc\'{i}a (CSIC), Apartado 3004, 18080 Granada, Spain}

\author[0000-0003-4166-2855]{Cody A. Carr}
\affiliation{Center for Cosmology and Computational Astrophysics, Institute for Advanced Study in Physics \\ Zhejiang University, Hangzhou 310058,  China}
\affiliation{Institute of Astronomy, School of Physics, Zhejiang University, Hangzhou 310058, China}

\author[0000-0002-6085-5073]{Floriane Leclercq}
\affiliation{CNRS, Centre de Recherche Astrophysique de Lyon UMR5574, Univ Lyon, Univ Lyon1, Ens de Lyon, F-69230 Saint-Genis-Laval, France}
 
\author[0000-0001-7144-7182]{Daniel Schaerer}
\affiliation{Observatoire de Gen\`eve, Universit\'e de Gen\`eve, Chemin Pegasi 51, 1290 Versoix, Switzerland}
\affiliation{CNRS, IRAP, 14 Avenue E. Belin, 31400 Toulouse, France}

\author[0000-0002-2838-9033]{Aaron Smith}
\affiliation{Department of Physics, The University of Texas at Dallas, Richardson, TX 75080, USA}

\author[0000-0003-0470-8754]{Jens Melinder}
\affiliation{Department of Astronomy, Oskar Klein Centre, Stockholm University,106 91 Stockholm, Sweden}

\author[0000-0002-5808-1320]{M. S. Oey}
\affiliation{University of Michigan, Department of Astronomy, 1085 S. University Ave, Ann Arbor, MI 48109, USA}

 \author[0000-0002-5269-6527]{Swara Ravindranath}
\affiliation{Astrophysics Science Division, NASA Goddard Space Flight Center, 8800 Greenbelt Road, Greenbelt, MD 20771, USA}
\affiliation{Center for Research and Exploration in Space Science and Technology II, Department of Physics, Catholic University of America, 620 Michigan Ave N.E., Washington DC 20064, USA}

\author[0000-0001-7016-5220]{Michael Rutkowski}
\affiliation{Minnesota State University-Mankato, Dept. of Physics and Astronomy, TN141, Mankato, MN 56001, USA}

\author[0000-0001-9269-5046]{Bingjie Wang}
\thanks{NHFP Hubble Fellow}
\affiliation{Department of Astrophysical Sciences, Princeton University, Princeton, NJ 08544, USA}


\begin{abstract}
Characterizing the mechanisms and galaxy properties conducive to the escape of ionizing (LyC) emission is necessary to accurately model the Epoch of Reionization, and identify the sources that powered it. 
Using Hubble Space Telescope data, the Lyman-alpha and Continuum Origins Survey (LaCOS) is the first program to obtain uniform, multi-wavelength sub-kpc imaging for a large sample (42) of galaxies observed in LyC and enable statistically robust studies between LyC and resolved galaxy properties. 
Here, we characterize the morphology and galaxy merger properties of LaCOS galaxies and investigate their connection with the escape fraction of LyC emission \fesc.
We find strong anti-correlations between \fesc\ and size ($r_{20}$, $r_{50}$, and $r_{80}$) measured in filters containing emission from star-forming regions, and with the asymmetry and clumpiness in F150LP, a filter tracing UV continuum and \lya. 
We find that $\geq48\%$ of LaCOS galaxies, and $\geq41\%$ of LaCOS LyC-emitters are visually classified as galaxy mergers. Galaxies robustly identified as mergers in LaCOS are at advanced stages of interaction, close to coalescence. The \fesc\ properties of robust mergers and low-probability mergers cannot be differentiated statistically, and we only find significant difference between the two populations in terms of their sizes and LyC luminosity, robust mergers having larger values. We conclude that \fesc\ tends to be larger in galaxies with a small number of compact, centrally-located, UV-emitting star-forming regions, that mergers at advanced stages of interaction represent a sizable fraction of LyC-emitting samples at $z\sim0.3$, and that they can facilitate the escape of LyC photons from galaxies. 

\end{abstract}



\section{Introduction} \label{sec:intro}
The escape fraction of ionizing radiation \fesc\ is a key parameter needed to model the Epoch of Reionization ($z\gtrsim5.5$), an important phase transition of the Universe when primordial astrophysical objects ionized the intergalactic medium \citep[e.g.][]{Robertson2022}. The escape fraction is defined as the ratio between the flux of hydrogen-ionizing (Lyman Continuum, LyC, $\lambda<912\,$\AA) photons that escape the interstellar and circumgalactic media of a given source, and that which is intrinsically produced by that source. Assumptions on the \fesc\ of high-z sources can completely change the history of reionization \citep{Finkelstein2019,Naidu2020,Lin2024}: obtaining accurate constraints for this parameter is thus necessary to properly model the high-z Universe. Due to the absorption of LyC photons by the remaining neutral fraction of the intergalactic medium, it is essentially impossible to detect LyC at $z>4$ \citep{Inoue2014}. In order to estimate the \fesc\ during the Epoch of Reionization, one must use indirect methods which include calibrations obtained at lower redshifts \citep[see review in][]{Jaskot2025}, Bayesian inference  \citep{Begley2022,Begley2024,Kreilgaard2024}, or SED-fitting coupled with LyC escape prescriptions \citep{Papovich2025,Giovinazzo2025}. 

The two main astrophysical sources thought to participate in the reionization of the Universe are star-forming galaxies and Active Galactic Nuclei (AGNs). Star-forming galaxies emit LyC photons through their O and B stars, that efficiently produce LyC photons, and where strong stellar feedback helps clear the interstellar medium and facilitate LyC escape \cite[][Carr et al., in prep.]{Heckman2011,Trebitsch2017,Chisholm2017,Amorin2024,Flury2024,Carr2025,Komarova2025,Martin2024,Flury2025}. AGNs produce LyC through heating onto their accretion disks, and are thought to have large \fesc\ ($\sim$ 75\%) \citep{Cristiani2016,Grazian2018}, although see e.g. \cite{Micheva2017,Smith2024}. However, whether their number density and ionizing output at high-z are sufficient to reionize the Universe is still under debate \citep{Grazian2018,Hassan2018,Dayal2020,Matsuoka2023,Trebitsch2023,Harikane2023b,Yeh2023,Madau2024,Grazian2024}, making star-forming galaxies the candidates currently favored by the community \citep{Hassan2018,Dayal2020,Rosdahl2022,Trebitsch2022,Trebitsch2023,Matsuoka2023,Yeh2023}. Observations of LyC-emitting galaxies (LCEs) at intermediate and low redshift have provided important insights into the properties of galaxies that enable LyC escape. The role of both global \citep{Leitet2013,Izotov2016b,Izotov2021,Izotov2018b,Izotov2022,Flury2022b,Saldana-Lopez2022,Chisholm2022,Roy2024,
Wang2019, Izotov2021, Xu2022,Vanzella2018,Mostardi2015, Steidel2018, Fletcher2019, Nakajima2020,  Ji2020, Marques-Chaves2021, Pahl2021,Saxena2022a,Saxena2022b,Wang2025} and sub-kpc scales \citep{Kim2023,Marques-Chaves2024,Komarova2024,LeReste2025,Saldana-Lopez2025,Ji2025} galaxy properties into \fesc\ have been investigated, allowing for calibrations between galaxy properties and \fesc\ to be derived \citep{Izotov2018b,Chisholm2018,Chisholm2022,Flury2022b,Xu2022,Choustikov2024,Jaskot2024a,Mascia2024,Leclercq2024,Saldana-Lopez2025}. It is generally understood that LyC-emitting galaxies tend to have large specific star formation rates, high ionization states, produce stars in compact regions, and have low \hi\ coverage on the line of sight to their LyC production sites \citep{Jaskot2025}. Several mechanisms conducive to LyC escape from galaxies have been put forward in the literature, including suppressed feedback \citep{Jaskot2017,Carr2025}, ionizing feedback \citep{Gazagnes2018,Gazagnes2020,Flury2022b,Flury2025,Bait2024}, stellar and supernova feedback \citep{Chisholm2017,Amorin2024,Komarova2021,Komarova2024,Flury2025} and bursty star formation \citep{Trebitsch2017,Flury2025}. However, what triggers the formidable star-forming episodes in the extreme galaxies leaking LyC photons is still not well understood.

Galaxy mergers have been suggested by several studies as a mechanism that could facilitate LyC escape from galaxies through their impact on star formation and gas morphology \citep{Bridge2010,Bergvall2013,Purkayastha2022,LeReste2024,Yuan2024,Zhu2024,Maulick2024}. Galaxy mergers constitute $<5$\% of the general galaxy population at $z\sim0$ \citep{Allam2004,Lotz2008,Conselice2009,Darg2010}, but this fraction increases to $\sim10$\% for star-forming galaxies  \citep{Lotz2008,Kaviraj2015,Duncan2019} and up to $50$\% of the starburst population \citep{Robaina2009,Pearson2019b}. During gas-rich interactions, mergers repeatedly trigger starburst episodes following pericentric passages \citep{Faria2025}. Mergers tend to have significantly larger star-formation rates than isolated galaxies with otherwise similar properties, especially at later stages of interaction \citep{Patton2013, Stierwalt2015, Ferreira2025}. By creating numerous massive stars, starbursts naturally increase the intrinsic production of LyC photons, as well as boost their escape through stellar and supernova-driven feedback \citep{Trebitsch2017,Barrow2020,Ma2020, Choustikov2024}. Additionally, tidal forces in mergers drive part of the gas away from the center of galaxies \citep{Pearson2016}, thus enhancing \fesc\ in lines-of-sights oriented away from tidal tails \citep{LeReste2024,Ejdetjarn2025}. However, while galaxy interactions could theoretically help increase LyC production and escape, the existence and extent of this increase is not well characterized. In particular, there are currently very few constraints on the merger fraction in samples of confirmed LyC-emitters, and how they compare to the general and star-forming galaxy populations. Furthermore, no constraints on the interaction timescales characteristic of LyC-emitting mergers are currently available. Many galaxy properties, including e.g. the star-formation rate, metallicity or AGN excess strongly change as a function of merger progression, with the strongest differences with control galaxies generally observed a few 100 Myr around coalescence \citep{Pearson2025,Ferreira2025,Ellison2025,Faria2025}. Similarly, LyC escape may only be facilitated in specific stages of a galaxy merger, with implication for the observability of LyC emission during mergers.

Simulations of $z=5-10$ galaxies using simple prescriptions for \fesc\ have shown promising results, indicating galaxy mergers could indeed enhance LyC escape \citep{Kostyuk2024}. Yet, observational studies on the role of galaxy mergers in LyC-emitters ($z\lesssim 3$) have so far remained limited to small samples \citep[]{LeReste2024,Maulick2024}, or to samples where the identification of mergers, and sometimes, LyC-emitters, is challenging due to the high redshift ($z\geq$3) of sources \citep{Zhu2024,Mascia2025}. Finally, a few studies have reported LyC emission offset from centers of galaxies, and have proposed merger interactions as a possible explanation \citep{Yuan2024,Gupta2024}. However, the lack of continuum emission redward of the Lyman limit in the regions where the offset emission is detected warrants careful consideration, as it raises questions about the nature and origin of the observed LyC signal. 

The most commonly employed methods for merger identification are close-pair detection and morphological analysis. However, several factors conspire to make the accuracy of those methods poor as redshift increases. In particular, for morphological characterization, the low-surface brightness features enabling the detection of advanced mergers fade rapidly with redshift ($\propto (1+z)^4$), and galaxies become clumpier as a function of redshift \citep{Guo2015}, resulting in increased confusion between mergers and star-forming galaxies at $z>2$ \citep{Abruzzo2018}. For close pair analysis, lower mass companions may not be detected at high redshift, so that merger identification fails to detect robustly all but the highest mass, highest luminosity mergers. Therefore, studies in the $z<2$ Universe, where merger identification is the most accurate and the direct observation of LyC is possible, are needed to help constrain the role of mergers onto LyC emission and escape.

The Lyman alpha and Continuum Origins Survey (LaCOS) is a program that imaged 42 nearby ($z\sim0.3$) star-forming galaxies observed in LyC with the Hubble Space Telescope \citep{LeReste2025}. The goal of this survey is to identify the resolved galaxy properties linked to LyC emission and escape. \cite{LeReste2025} showed that that \fesc\ correlates best with properties evaluated in the small apertures centered on the brightest UV clumps. This suggests that a small number of bright, unobscured UV-emitting clusters are responsible for the bulk of ionizing photons escaping on our line-of-sight. \citet{Saldana-Lopez2025} also found that LyC-emitters tend to have compact \lya\ haloes, indicating a higher fraction of \lya\ photons escape directly from HII regions in strong LyC-emitters. Owing to the availability of relatively deep, high-resolution imaging across several bands and the sufficiently low redshift of the sample, LaCOS is uniquely suited to in-depth morphological studies and merger identification of LyC-emitting galaxies. In this paper, we derive parameters characterizing the morphology of galaxies in the rest-frame UV and optical bands, to estimate if the morphology of galaxies across wavelengths can be used to estimate \fesc\ and get insights on the processes responsible for LyC escape. Additionally, we visually identify mergers and characterize the advancement of interactions, to assess if galaxy mergers constitute a large fraction of LyC-emitting galaxies and get insights on the timescales where LyC escape occurs in mergers.

This paper is structured as follows. In section \ref{sec:data} we describe the data and methods used to derive morphological parameters and merger properties. In section \ref{sec:results-morpho}, we present the morphometrics for our sample, place them in the context of the $z\sim0.3$ galaxy population, and evaluate trends with \fesc. In section \ref{sec:results-mergers} we show the results of the merger characterization and estimate the impact of merger interactions on LyC properties. Finally, we discuss the results in section \ref{sec:discussion} and present a summary and conclusion in section \ref{sec:conclusion}.

Throughout this work, we assume a standard flat $\Lambda$CDM cosmology with $H_0=70\,\rm{km}\,\rm{s}^{-1}\,\rm{Mpc}^{-1}$ and $\Omega_m=0.3$.

\section{Data and methods} \label{sec:data}
\subsection{LaCOS HST photometry}
LaCOS \citep{LeReste2025,Saldana-Lopez2025} is a cycle 30 HST program that imaged 42 nearby ($z\sim0.3$) star-forming galaxies observed in LyC as part of the Low-z Lyman Continuum Survey and its extension including archival observations \citep[LzLCS+][]{Flury2022a}. The LzLCS is a flagship survey that observed 67 low-z galaxies in LyC and significantly increased the number of known LyC-emitting objects as well as extended the galaxy property parameter space for these objects \citep{Flury2022b}. LaCOS builds upon LzLCS+ by delivering high-resolution, multi-wavelength imaging for a subset of galaxies representative of the full survey, aimed at investigating the link between resolved galaxy properties and LyC emission and escape. About half (22) of the galaxies observed in LaCOS are confirmed LyC-emitters with \fesc=$0.01 - 0.49$, while the rest are either not detected or have low-significance detections ($<2\sigma$) with \fesc$<0.02$. The availability of non-detections is crucial, since those galaxies serve as a control sample with the exact same set of observations and selection criteria as the LyC-emitters. LaCOS observations consist of imaging in 5 filters, including two rest-frame UV filters (F150LP, F165LP) and three rest-frame optical filters  (F438W, F547M, F850LP). With a $\sim0.1"$ PSF ($<400$ pc physical scale) and the rich multiwavelength dataset available, it is uniquely suited to morphological studies of LyC emitters. We refer the readers to \cite{LeReste2025} for a description of the LaCOS data reduction. Below, we describe the methods used to measure the morphology of the galaxies and identify galaxy mergers in the sample.

\subsection{Morphometrics} \label{sec:methods-morph}
Non-parametric morphological estimators, also sometimes called morphometrics, are parameters characterizing the distribution of light in galaxies, that do not require assumptions on the shape of their light profile. A large number of such parameters have been developed, among which the concentration $C$\footnote{Initially proposed as a visually-defined parameter \citep{Morgan1958,Kent1985}.} \citep{Abraham1994,Abraham1996, Bershady2000}, the asymmetry $A$ \citep{Elmegreen1992,Abraham1996,Conselice2000,Conselice2003}, the clumpiness $S$ \citep{Conselice2003}, the Gini coefficient $G$ \citep{Abraham2003, Lotz2004}, and the second-order moment of brightest 20\% of light $M_{20}$ \citep{Lotz2004} are the most well-known, though not an exhaustive list \citep[see also][]{Freeman2013,Wen2014,Pawlik2016}.

To explore possible correlations between \fesc\ and the morphology of galaxies, we derive several morphometrics for LaCOS galaxies in all filters available, covering rest-frame wavelengths $\lambda\sim1100-8500\,$\AA. Specifically, we derive $A$, $C$, $S$, $G$, $M_{20}$, the shape asymmetry $A_s$ \citep{Pawlik2016}, and the radii containing 20\%, 50\% and 80\% of the light, $r_{20}$, $r_{50}$ and $r_{80}$ using the commonly used package \texttt{statmorph} \citep{Rodriguez-Gomez2019}. We note that radii were already derived for LaCOS galaxies UV filters in \cite{Saldana-Lopez2025}, but using a different method based on integration in circular radii. Prior to deriving the morphological parameters, the images are background-subtracted using sigma-clipped median filtering and a 2D background model with the \texttt{Background2D} functions in \texttt{photutils}. Segmentation maps are generated with the \texttt{photutils} \texttt{detect\_sources} routine, using a $1\sigma$ threshold to detect sources with a minimum of 8 pixels. The apparent boundaries of the galaxies are consistently identified in F547M filter images ; for consistency, this filter was used to define the segmentation map for each galaxy. We note that the exact choice of the filter used to define segmentation maps has little impact on the morphometrics in \texttt{statmorph}. Indeed, the user-defined segmentation map is used to calculate noise properties, and different segmentation maps adapted to the calculation of specific parameters are derived by \texttt{statmorph}. Morphometrics are calculated for all galaxies and all filters available in LaCOS using the \texttt{SourceMorphology} class from \texttt{statmorph} and using the segmentation map defined using the F547M filter. We do not measure morphometrics in continuum-subtracted \lya\ images produced from the combination of F150LP and F165LP and presented in \cite{LeReste2025}. The patchiness of these continuum-subtracted maps prevents accurate segmentation, and due to the presence of \lya\ absorption in several galaxies, morphometrics cannot be robustly derived using \texttt{statmorph}. Custom morphological parameters characterizing \lya\ emission were derived and compared to \fesc\ in \cite{Saldana-Lopez2025}, we refer the reader to this study for a comparison of \lya\ morphology and LyC emission.

\subsection{Merger identification}
\label{sec:methods-merg}
The first step to estimating the impact of mergers on LyC emission is to identify merging galaxies as accurately as possible in LaCOS.
Galaxy merger identification typically either proceeds via close-pair searches or via morphological identification. The first method identifies galaxies in close proximity that have a high chance of merging in the future, searching for objects with small projected separations and relative velocity separations \citep{Patton2000,Ventou2019}. Morphological classification instead targets galaxies at relatively advanced stages of interaction, taking advantage of the fact that tidal gravitational forces significantly perturb the stars and interstellar medium of the galaxies involved in a merger \citep{Holmberg1941,Toomre1972,Conselice2000,Lotz2004}. The identification of galaxy mergers through their morphology can be done either visually \citep{LeFevre2000,Darg2010,Kartaltepe2015}, using non-parametric morphological parameters \citep{Conselice2003,Lotz2004,Scarlata2007,Holwerda2011,Wen2014,Pawlik2016}, or using machine learning \citep{Ferreira2020,Walmsley2020,Pearson2019}. The later two methods require calibration, which can be done by using visual identification as a standard \citep{Conselice2003,Lotz2004,Lotz2008,Walmsley2020}, or with mock images from simulations, where the merger history is known \citep{Pearson2019,Ferreira2024}.

All of these methods have drawbacks and advantages, so that the choice of an identification scheme strongly depends on the data on which it is performed. Close pair searches require both objects in the pair to have accurate redshifts, and is therefore typically limited to brighter objects in large-scale surveys. Additionally, this method does not allow for the identification of mergers at advanced stages, when the galaxies in a pair are too close to be resolved, or have already merged. Visual identification, while being the oldest of all methods, is still often employed as the human eye is uniquely suited to identifying subtle morphological patterns in the light profile of galaxies, and is the simplest method to implement for small sample sizes. However, it can be relatively biased, requiring a sufficiently large number of classifiers to yield accurate results, and requires crowdsourcing in the case of large samples \citep{Lintott2008,Darg2010}.
Morphometrics, on the other hand are extremely convenient when exploring large datasets, as non-parametric morphological parameters can be derived in bulk. However, the validity of the criteria used to select mergers via morphometrics are 1) strongly dependent on the surveys they are derived from (such as the redshift and mass ranges considered, see e.g. \cite{Abruzzo2018}), and 2) are often calibrated using visual identification and thus, ultimately suffer from the same issues as visual identification. In particular, survey-to-survey signal-to-noise and wavelength coverage variations can strongly impact the validity of certain criteria \citep{Lotz2008,Mager2018}. Finally, machine learning (ML)-based approaches offer a new avenue for identifying mergers beyond non-parametric morphological estimators, by learning complex patterns in high-dimensional parameter spaces. In recent years, the training of ML algorithms has been performed on mock observations from simulations, alleviating the bias that could potentially be introduced by visual classification \citep{Pearson2019,Ferreira2024}. However,  similarly to morphometrics, the specificity of the training set and image quality can impact the adaptability and applicability of the models to real data \citep{DominguezSanchez2023,Bickley2024}.

\begin{figure*}[t]
    \centering
    \includegraphics[width=\linewidth]{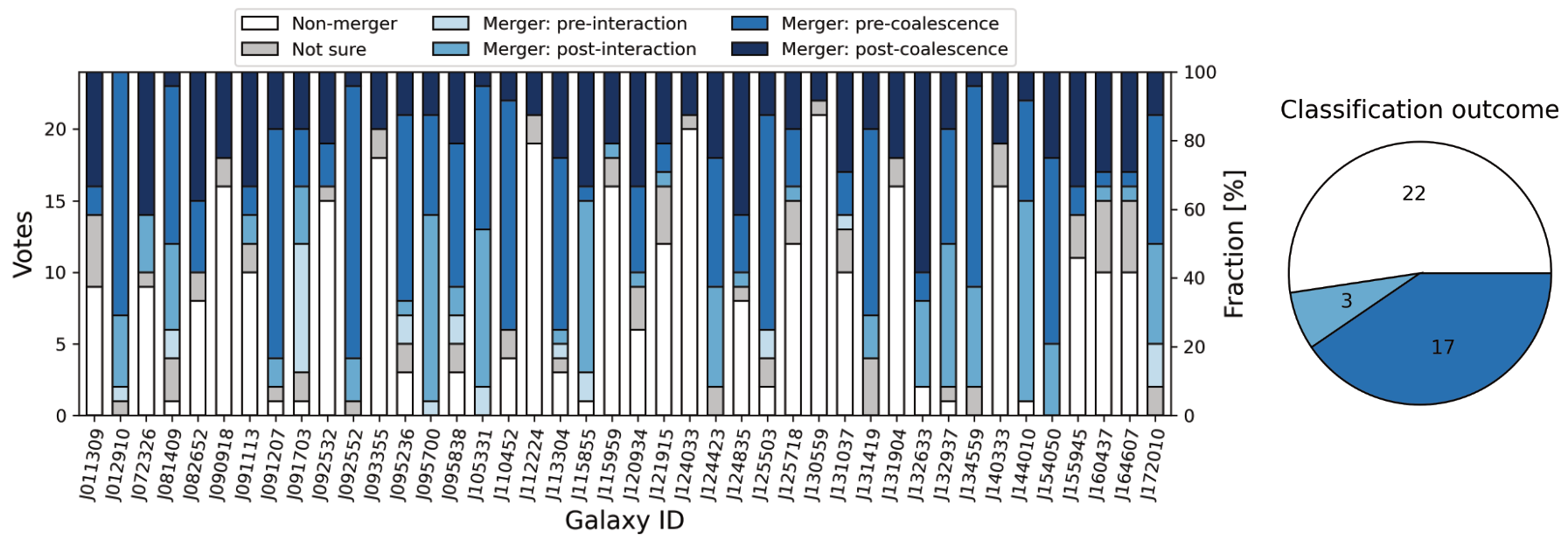}
    \caption{Left panel : distribution of visual merger classification votes for LaCOS galaxies. Each galaxy was classified by 24 of the LaCOS collaboration members. There is considerable variation in the votes for any given galaxy, illustrating the importance of a large number of classifiers to obtain a consensus answer. Right panel : Pie chart showing the number of galaxies per category resulting from the classification.}
    \label{fig:vote}
\end{figure*}

LaCOS contains only 42 galaxies at $z=0.22-0.32$, with specific galaxy property and imaging parameter spaces \citep{LeReste2025}. These make both morphometrics criteria developed for general galaxy surveys and the use of existing CNNs trained on survey-specific mock images derived from simulations inadequate. While all of the galaxies have SDSS spectroscopy, they tend to have low stellar masses (with median $M_*=10^9\,M_\odot$, and as low as $1.8\times10^8\,M_\odot$), meaning that possible lower-mass companions may not have available spectroscopic data. Additionally, we expect mergers at later stages of interaction to be important for LyC emission, as this is when the largest enhancements in SFRs are found \citep{Ferreira2025}, and simulations looking at the post-merger \fesc\ evolution find the largest \fesc\ at coalescence \citep{Kostyuk2024}. Thus, we choose to proceed with visual identification for our sample of 42 galaxies. To mitigate bias that could be introduced by visual identification, all members of the LaCOS collaboration were asked to perform the classification. In total, each galaxy was classified by $24$ people.

Visual classification participants were shown image panels showing optical color-composite RGB (with B=F438W, G=F547M and R=F850LP), and single filter images of all galaxies, with both large-scale (24" on the side, corresponding to $\sim100\,$kpc) and 20 kpc cutout images for color-composites, as shown in Figure \ref{fig:mergclass_thumb} in Appendix \ref{app:visual_class_panel}. Importantly, the color-composites are made by normalizing the flux in each filter to its own maximum, to highlight structural features across all bands. LaCOS galaxies are highly star-forming, thus traditional RGB images using optical filters scaled to a common maximum are fully dominated by emission in the bluer bands, eclipsing faint isophotes from redder bands. In many cases, the morphological features that allow for the classification of LaCOS objects as mergers are mostly visible in the redder optical band.

In addition to classifying the galaxies in merger and non-merger categories, the classifiers were asked to visually asses the merger stage to obtain a measure of the advancement of the merger. To do so, we used broad categories split between pre-interaction and post-coalescence (following Mantha et al., in prep. and similar to the approaches in e.g. \cite{Veilleux2002} and \cite{Pan2019}). Each merger category was assigned a score $s_i$ ranging from 1 to 4 tracking the advancement of the merger. The specific options classifiers could choose, and associated merger stage scores were:
\begin{enumerate}
    \item \textbf{Not a merger:} no score.
    \item \textbf{Pre-interaction:} $s_i=1$. Well-separated galaxies that do not show any morphological distortion, i.e., incoming pairs, before the first pericenter passage.
    \item \textbf{Post-interaction:} $s_i=2$. Well-separated galaxies, but showing morphological distortion.
    \item \textbf{Pre-coalescence:} $s_i=3$. Two (or more) galaxies embedded within each-other’s light envelopes and displaying disturbed morphological features.
    \item \textbf{Post-coalescence:} $s_i=4$. The galaxies' central core regions have merged into one or are visibly indistinguishable, but the system has disturbed morphological features (e.g. tidal tails, faint asymmetric components, shells).
    \item \textbf{Not sure:} no score.
\end{enumerate}
We note that although the classification scheme outlined above divides galaxy mergers into distinct visual categories based on morphological features, there is overlap and potential ambiguity between these categories, which do not exactly map onto specific merger timescales. Nevertheless, these classifications still provide a useful measure of the overall progression of a merger.

Figure \ref{fig:vote} displays the distribution of votes resulting from the visual classification, demonstrating significant dispersion in classification outcomes. For this reason, and to select solely for mergers and limit the inclusion of non-merging irregular galaxies that could potentially be present in the sample, we establish a strict criterion for identifying a galaxy as a merger. We take into account all of the votes towards any merger category (with scores $s_i=1-4$) and establish a threshold based on the fraction of votes going to any merger category $P_{merg}=\frac{N_{merg}}{N_{tot}}$. For each galaxy, the merger identification process through voting can be approximated by a series of Bernoulli trials with yes/no answers. To set a baseline for classification, we therefore consider a binomial distribution, with random classification yielding a 50\% chance that a galaxy is classified as a merger for each vote. We use the \texttt{histogram} package developed in \citep{Flury2022a} to calculate the 3$\sigma$ confidence interval for a binomial distribution with 24 choose 12 assuming maximal variance in each vote. This corresponds to a merger vote fraction $P_{merg}=79\%$. We therefore select all galaxies with $P_{merg}\geq79\%$ as robust galaxy mergers. In Appendix \ref{app:morph_merg_comp} we compare the visual classification to traditional classifications using morphometrics \citep{Conselice2003,Lotz2008,Pawlik2016}. To obtain a broad measure of the advancement of a merger, we calculate the merger stage as $s_{merg}=\frac{\sum_i s_i}{N_{merg}}$, the average of scores for votes towards merger classifications. 

A table containing the morphometrics and merger parameters derived here can be found in the electronic version of this manuscript. A representative set of parameters are described in Appendix \ref{app:table} Table \ref{tab:morph}.

\section{Morphology of the LaCOS galaxies}  \label{sec:results-morpho}
\subsection{Correlation with the escape fraction}
 Here, we explore and characterize possible correlations between galaxy morphology and \fesc\, using morphometrics measured at different wavelengths in LaCOS filters. To that aim, we calculate the Kendall $\tau$ rank correlation coefficient \citep{Kendall1938} between \fesc\ and the various morphometrics derived in this work. The Kendall $\tau$ coefficient  measures the association between two variables through ordinal association, and thus provides an estimate of the strength of a correlation. Unlike the Pearson coefficient \citep{Pearson1895} which assumes linearity and normally distributed data, the Kendall coefficient is non-parametric and makes no assumption on data distribution. To enable the proper treatment of upper limits, the Kendall $\tau$ coefficients and associated $p$-values are calculated using an approach allowing for the inclusion of censored data \citep{Isobe1986}. Specifically, we use a publicly available python library using code adapted from \cite{Flury2022b} and presented in \cite{Herenz2025}. Following the convention in other LaCOS manuscripts \citep{LeReste2025,Saldana-Lopez2025}, we deem a correlation ($\tau>0$) or anti-correlation ($\tau<0$) robust when $p<1.35\times10^{-3}$ and tentative when $1.35\times10^{-3}<p<0.05$. These respectively correspond to a $>3\sigma$ and $2-3\sigma$ confidence that the null hypothesis, stating a coefficient of this magnitude or higher could have arisen purely by chance, can be rejected.
 
 \begin{figure*}[t]
    \centering
    \includegraphics[width=0.9\linewidth]{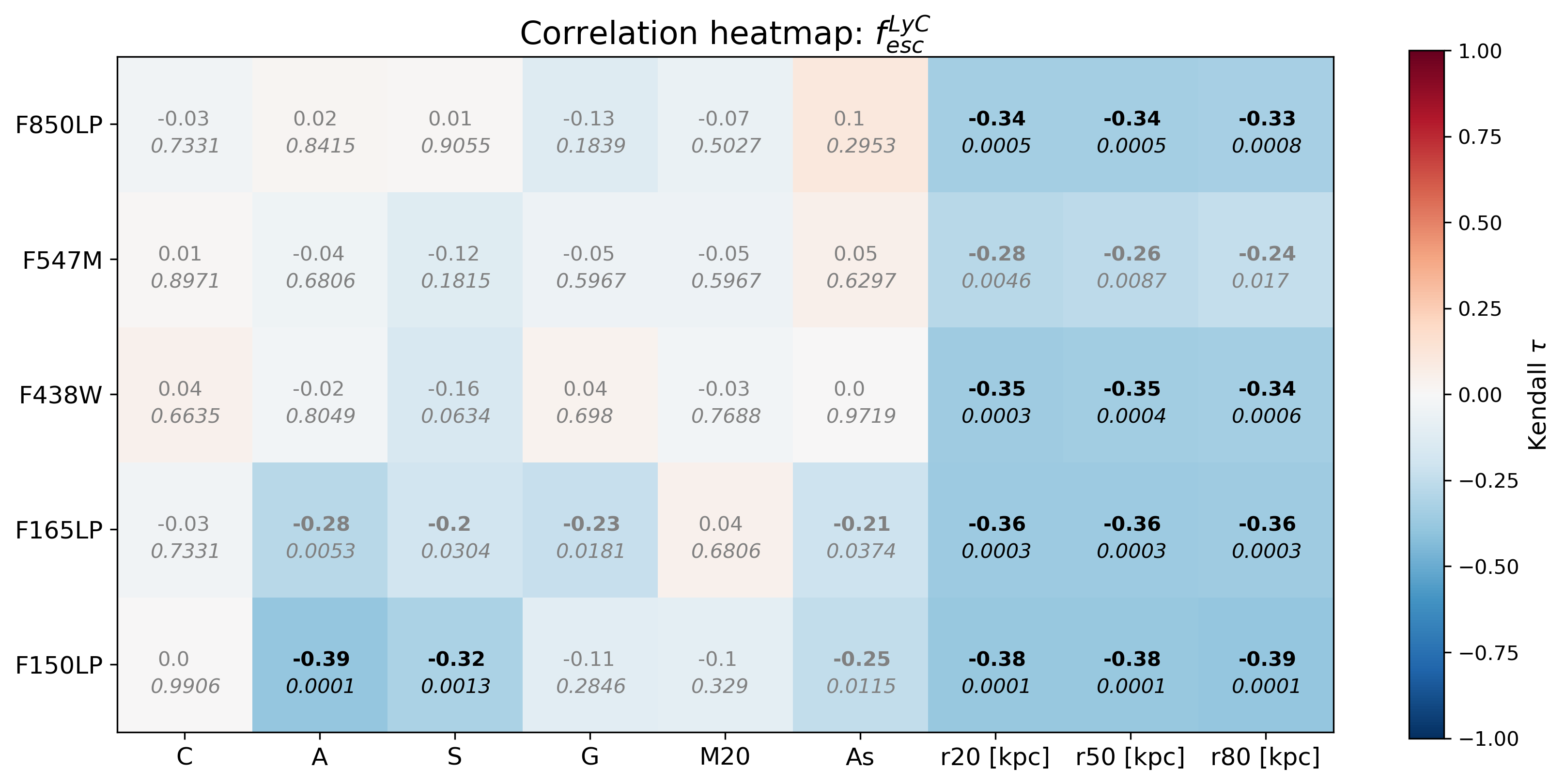}
    \caption{Heatmap showing the Kendall $\tau$ correlation coefficient and associated p-value between non-parametric morphological estimators and the \fesc\ for LaCOS galaxies in different filters. The Kendall $\tau$ values are shown, with corresponding p-values displayed directly below in italic. Robust correlations ($p<1.35\times10^{-3}$) are indicated in bold  black text, tentative correlations ($0.05<p\leq1.35\times10^{-3}$) in bold gray text and non-significant correlations ($p\geq0.05$) in regular gray text.}
    \label{fig:heatmap_fescLyC}
\end{figure*}

 In Figure \ref{fig:heatmap_fescLyC}, we present a heatmap showing the Kendall $\tau$ coefficient and associated $p$-values between morphometrics and \fesc\ in all available LaCOS HST filters. For robust correlations, the Kendall $\tau$ value is shown in black bold fonts, for tentative correlations it is shown in gray bold font. Non-significant correlations are shown in normal gray fonts. There is, for most filters, a relatively high degree of anti-correlation ($\tau<-0.33$, $p<8\times 10^{-4}$) between \fesc\ and the size of the galaxies as measured using $r_{20}$, $r_{50}$ and $r_{80}$. The only exception is for filter F547M, that only shows a tentative anti-correlation between galaxy sizes and \fesc. This may be explained by the fact that the filters where an anti-correlation can be found tend to trace star-forming regions. Indeed, F150LP and F165LP are both rest-frame UV filters, and F438W traces emission at rest-frame blue optical wavelengths, therefore containing emission from recently formed massive stars. F850LP, while sampling rest-frame red optical wavelengths, does include \ha\ emission at the edge of the bandpass for 29 of the galaxies in LaCOS (with $z\gtrsim0.27$), also tracing recent star-formation. Isolating galaxies with \ha\ contribution in the F850LP bandpass, we get a significantly stronger anti-correlation between $r_{50}$ and \fesc\ ($\tau=-0.43$, $p=4.5\times 10^{-4}$). Accordingly, for galaxies without \ha\ in the F850LP bandpass, we do not find a statistically significant anti-correlation between \fesc\ and $r_{50}$ ($\tau=-0.21$,$p=0.26$).
 This indicates that galaxies with small sizes and specifically, small star-forming regions, tend to have elevated escape fractions. A link between small UV-emitting regions and \fesc\ has been shown by several previous studies \citep{Kim2023,Flury2022b,LeReste2025}. Here we show that this may also be found in optical filters covering emission associated with recent star-formation.

Another strong anti-correlation is seen between the asymmetry in the F150LP filter and \fesc. The F165LP filter also shows a tentative anti-correlation between asymmetry and \fesc. Asymmetry in the rest-frame optical has been used as a tracer of galaxy interaction \citep[e.g.][]{Conselice2000,Conselice2003}, but here we find no association between asymmetry measured in the optical and \fesc. UV emission solely traces recent star-forming regions, while the redder stellar continuum traces older stars. As such, the asymmetry parameter in the UV can be seen as a proxy for the spatial distribution and number of star-forming clumps. Therefore, the anti-correlation between the F150LP asymmetry and \fesc\ likely indicates that galaxies with a small number of centrally-located UV clumps are more likely to have large \fesc. This is also reflected in the anti-correlation between \fesc\ and the clumpiness in F150LP, which measures the fraction of light in small scale structures, and \fesc. The stronger anti-correlations found in F150LP as compared to F165LP could be due to the fact that F150LP contains \lya\ emission, and that the sampled UV continuum is bluer, thus better tracing the regions emitting LyC. 

\cite{LeReste2025} found a strong correlation between \lya\ observables (specifically the luminosity and EW) measured in small apertures around the brightest UV emission region and \fesc\ on the line-of-sight. This was interpreted as the escaping LyC emission being contributed by a small number of bright, unobscured UV clumps. The anti-correlation between morphometrics in the F150LP filter, which contains \lya\ emission and stellar UV continuum, also supports this hypothesis. This is additionally supported by results in
\cite{Saldana-Lopez2025}, which measured the \lya\ halo properties of LaCOS galaxies. They found a strong anti-correlation between \fesc, $r_{20}$ and $r_{50}$ as measured in the \lya\ images ($\tau<-0.41$, $p<10^{-4}$), although we note their radii were derived using a different method. This is also in line with their results on the \lya\ halo fraction, which showed that the \lya\ cores, instead of the haloes, drive the correlation with \fesc. Finally, \cite{Saldana-Lopez2025} found a strong correlation between \fesc\ and $r_{90}$/$r_{20}$ in \lya, a measurement analogous to the concentration parameter as defined in \texttt{statmorph}. Here, we do not find any  correlation between the concentration parameter in F150LP and \fesc, which may be due to the filter containing both \lya\ and UV continuum emission, or the difference in methods employed to measure $C$ and radii. 

\begin{figure*}
    \centering
    \includegraphics[width=0.9\textwidth]{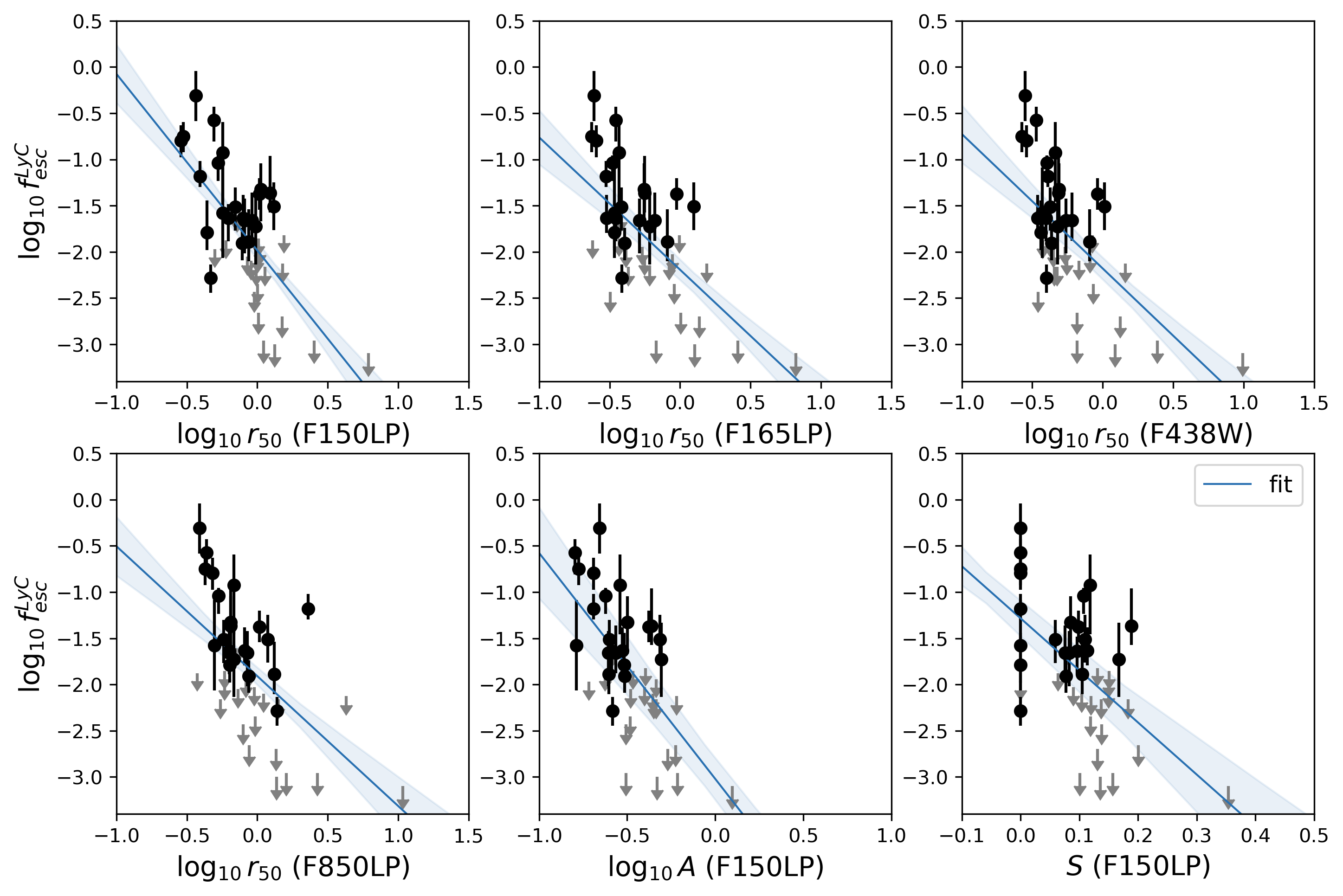}
    \caption{Fits between \fesc\ and the morphometrics found to be robustly anti-correlated with \fesc. }
    \label{fig:morph_fits}
\end{figure*}

We fit log-log and log-linear relations between $\log_{10}f_{esc}^{LyC}$ and the morphometric parameters found to significantly anti-correlate with \fesc\, as shown on Figure \ref{fig:morph_fits}. For that purpose, we use the python package \texttt{linmix} \citep{Kelly2007}, that performs linear fits to censored datasets using a Bayesian framework. This allows for the accurate treatment of upper limits in the fits to $\log_{10}f_{esc}^{LyC}$. Specifically, for each parameter $x$, we fit a function fo the form:
$$\log_{10}f_{esc}^{LyC}=a+b\cdot x$$
with fitting parameters $a$ and $b$ shown in Table \ref{tab:linmix_pars}. Either the log-log or log-linear fit is presented, depending on which yields the lower $\chi^2$ value.
As with previous correlations involving single variables and \fesc, we find a relatively large rms scatter around the fit of $\sim$0.5 dex. In comparison, the best multivariate models currently available have a characteristic rms scatter of $\sim$0.3 dex \citep{Jaskot2024a}.

\begin{table}[]
    \centering
    \caption{Fitting parameters and rms scatter around the fit for the log-log and log-linear relations between $\log_{10}f_{esc}^{LyC}$ and morphometrics $x$. The radii are in kpc.}
    \begin{tabular}{c|c|c|c}
    \hline
    $x$ & a & b & $rms$\\
    \hline
    $\log_{10}\,r_{50,\,F150LP}$ & $-1.98\pm0.08$ & $-1.9\pm0.31$ & 0.46 \\
    $\log_{10}\,r_{50,\,F165LP}$ & $-2.19\pm0.1$ & $-1.43\pm0.27$ & 0.48 \\
    $\log_{10}\,r_{50,\,F438W}$ & $-2.18\pm0.11$ & $-1.45\pm0.29$& 0.49 \\
    $\log_{10}\,r_{50,\,F850LP}$ & $-1.91\pm0.09$ &$-1.41\pm0.3$ & 0.51 \\
    $\log_{10}\,A_{F150LP}$ & $-3.02\pm0.22$ & $-2.45\pm0.43$ & 0.48 \\
    $S_{F150LP}$ &$-1.28\pm0.15$  & $-5.63\pm1.2$ & 0.52 \\
    \hline
    \end{tabular}
    \label{tab:linmix_pars}
\end{table}

\subsection{Comparison to the general galaxy populations}

\begin{figure*}
    \centering
    \includegraphics[width=0.9\textwidth]{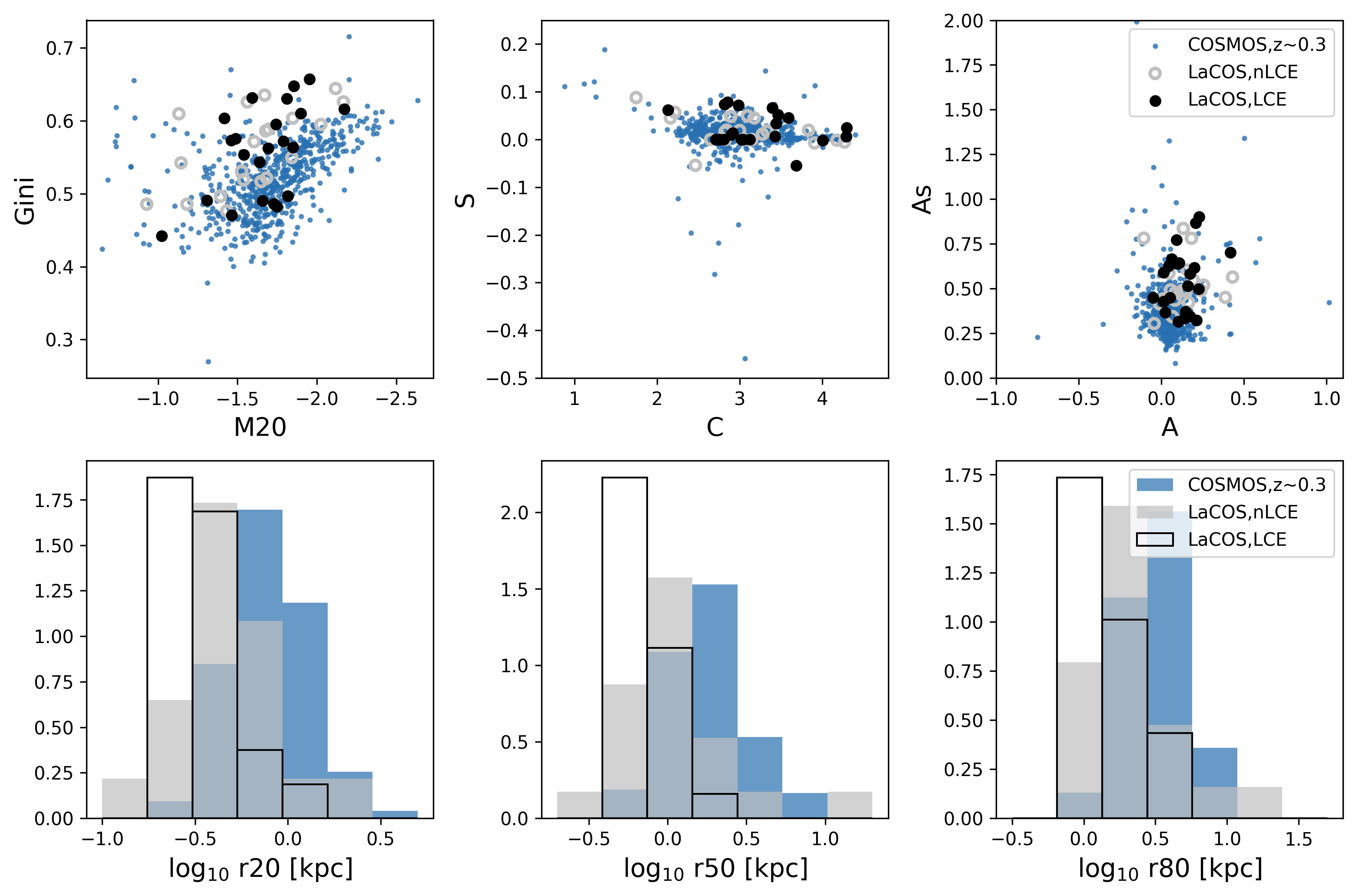}
    \caption{Comparison of the morphometrics parameter space for LaCOS galaxies as measured in the F850LP filter (black for LyC-emitters and gray for galaxies with upper limits) as compared to COSMOS galaxies with similar redshifts ($z\sim0.3$) and stellar masses ($7.8<\log_{10}\,\rm{M}<11.2$)}, with morphometrics measured in F814W (shown in blue). On the top panels, from left to right, we show the M20-Gini, C-S and A-As diagrams. On the bottom panels, from left to right we show density plots for $r_{20}$, $r_{50}$ and $r_{80}$.
    \label{fig:comp_cosmos}
\end{figure*}

We have explored correlations between \fesc\ and morphometrics in different filters in the previous section. In the following, we compare the morphometrics derived in LaCOS to those of the general galaxy population to get a sense of the parameter space occupied by LyC emitters and candidates.

We compare LaCOS morphometrics to those available for galaxies in the Cosmic Evolution Survey \citep[COSMOS]{Scoville2007}. The COSMOS survey is very well suited to such a comparison, as it is a wide, multi-wavelength survey of galaxies covering a large redshift range and having HST imaging. We note, however, that at the redshift range of interest, COSMOS galaxies do not have LyC observations. Therefore, we do not know if, or which of the COSMOS galaxies are LyC-emitters. Nevertheless LyC-emitters are rare at $z\sim0$, so we expect a very small number of LyC-emitters in COSMOS in the redshift range considered, if any.

To enable comparisons between LaCOS and COSMOS, we select COSMOS galaxies at the same redshifts as LaCOS objects ($z=0.22-0.32$) using the spectroscopic redshift information in \cite{Khostovan2025}, yielding an initial sample of 3573 galaxies. To limit selection effects, we apply an additional cut on the stellar mass using the catalog in \cite{Weaver2022} to select COSMOS galaxies within the LaCOS mass range ($7.8<\log_{10}\,\rm{M}<11.2$), reducing the COSMOS control sample to 637 galaxies. While morphometrics have been derived for COSMOS in \cite{Scarlata2007}, differences in methods between \texttt{statmorph} and the dedicated software developed for the purpose of analyzing COSMOS galaxies could cause systematic differences in morphometrics and bias the comparison. Therefore, we adopt an approach similar to that employed in \cite{Scarlata2007}, but using \texttt{statmorph} to compare COSMOS and LaCOS morphometrics. First, we use the \texttt{Python} \texttt{SExtractor} implementation \texttt{sep} \citep{BertinAnouts1996,Barbary2016} to derive segmentation maps and replace the pixels from sources in the field other than the main target with noise to create clean images. We then use the semiminor and semimajor axes $a$ and $b$ retrieved by \texttt{sep} to calculate the ellipticity $e=1-b/a$ and position angle for the source. Additionally, we use the source flux to define a "total radius" $R_{tot}$ through integration in circular apertures and extract stamps around the targets that are $3\times R_{tot}$ in size. We use the package \texttt{PetroFit} to calculate the Petrosian radius in the masked stamps \citep{Geda2022,petrofitZenodo}. Finally, we derive an initial segmentation map that is an ellipse of major-axis equal to the Petrosian radius and with ellipticity and position angles as derived using \texttt{sep}. We then feed the corresponding stamps and segmentation maps to \texttt{statmorph} for morphometrics calculation. For COSMOS galaxies \citep{Koekemoer2007,Massey2010}, we use the $0.03"$ pixel scale HST/ACS F814W mosaic images retrieved through the NASA/IPAC Infrared Science Archive \citep{10.26131/irsa178}. Morphometrics calculation failed for 25 galaxies due to contamination by neighboring stars and sources, incomplete HST coverage, or imaging artifacts. This results in 612 COSMOS galaxies with redshifts and stellar masses matched to LaCOS and valid morphometrics. We re-derive LaCOS morphometrics in the F850LP filter (close to F814W in terms of wavelength coverage) using the same method as that applied to COSMOS galaxies. We note that the morphometrics derived for the comparison to COSMOS images in LaCOS have values similar as when using the method described in \ref{sec:methods-morph}. The comparison of the morphometrics between LaCOS and COSMOS galaxies is shown in Figure \ref{fig:comp_cosmos}. 

Generally, and in agreement with the correlation coefficients shown in Figure \ref{fig:heatmap_fescLyC}, the LyC-emitters and non-emitters in LaCOS tend to occupy similar parameters spaces in $M_{20}$, Gini, A, A$_{s}$, C and S when measured in the optical. We run Kolmogorov-Smirnov (K-S) tests using the \texttt{scipy} function \texttt{kstest} for all morphometrics between the LaCOS and COSMOS samples, and find that the LaCOS galaxies have significantly larger Gini, $A$ and $A_s$ parameters and smaller $r_{20}$, $r_{50}$, and $r_{80}$ in the optical than the general galaxy population at $z\sim0.3$ (with $p<1.35\times10^{-3}$). This may be partly due to sample selection, as one of the criteria used to select LzLCS galaxies requires large $\Sigma SFR$, necessarily yielding galaxies with small radii and large star formation rates. However, and in accordance with results from \ref{sec:results-morpho}, LyC-emitters have the smallest sizes, as compared both to non LyC-emitters in LaCOS and the general galaxy population.
Apart from their radii, the LyC-emitters and non-emitters in LaCOS do share similar parameters space. We therefore conclude that when it comes to concentration, asymmetry, Gini and $M_{20}$, any difference in morphologies between LaCOS galaxies and COSMOS is likely due to the criteria used to select LyC candidates in LzLCS, and in particular requirement on $\Sigma SFR$ as outlined above. Interestingly, LaCOS galaxies are located in the part of the $M_{20}$-Gini parameter space where galaxy mergers tend to be found \citep{Lotz2008}. In the following, we explore the galaxy merger properties of LyC-emitters and non-emitters in LaCOS.

\section{Merger properties of LaCOS galaxies}
\label{sec:results-mergers}
\begin{figure*}[t]
    \centering
    \includegraphics[width=0.95\linewidth]{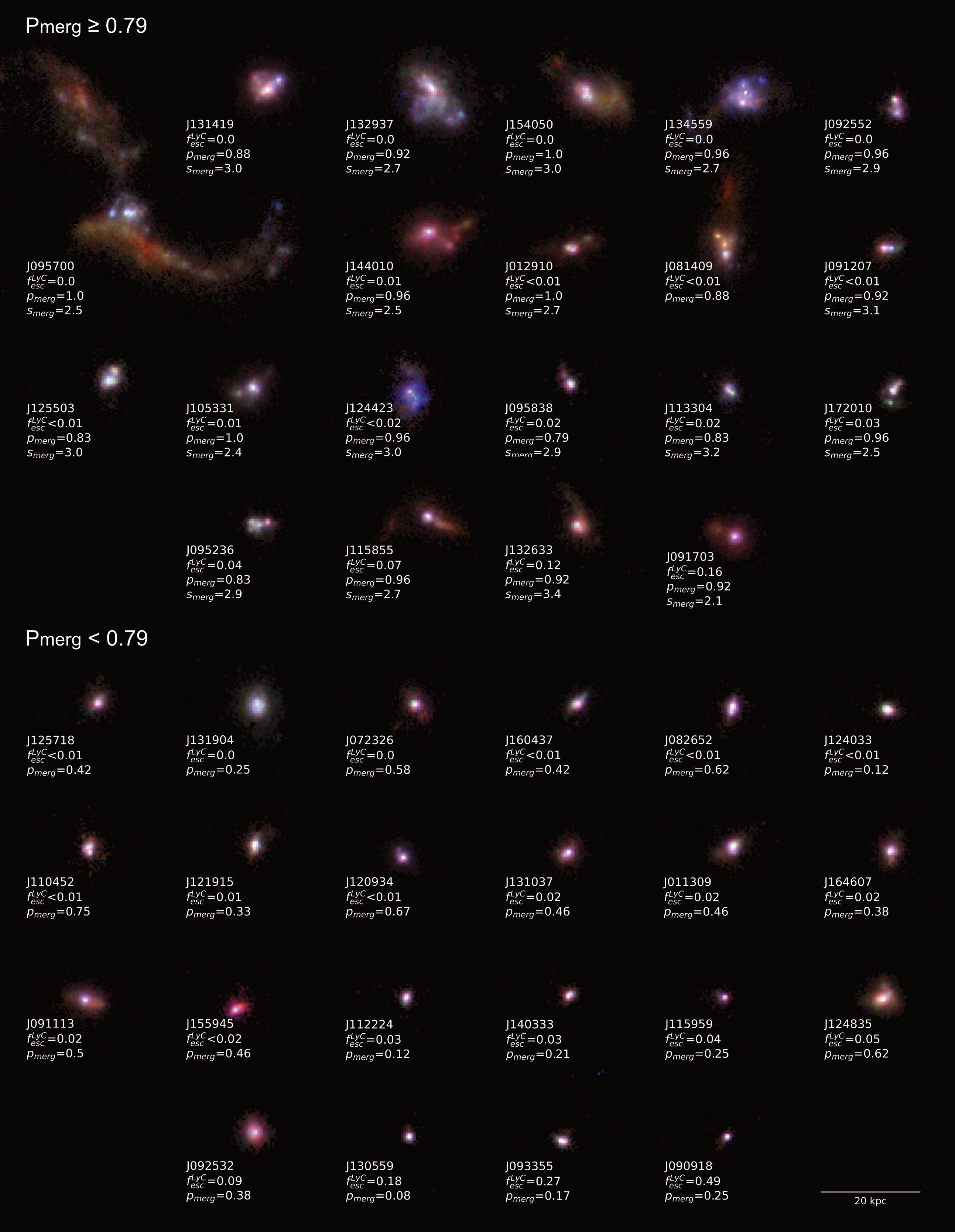}
    \caption{RGB images of LaCOS galaxies, split according to the fraction of merger votes. The top most panels show galaxies with $P_{merg}\geq79\%$, robustly identified as mergers ($\geq3\sigma$), while the galaxies on the bottom panels show galaxies with low fraction of merger votes. In these categories, galaxies are ordered from top to bottom by \fesc.} Next to each galaxy, we show the values for \fesc, fraction of merger votes, and the merger stage. Galaxies are organized by increasing \fesc\ from top to bottom and left to right for the robust mergers and the galaxies with low merger vote fractions. A bar on the lower bottom right corner indicates the 20 kpc scale.
    \label{fig:rgbs}
\end{figure*}

 Here, we describe the results of the merger identification and stage evaluation as described in section \ref{sec:methods-merg}. Specifically, we describe the merger fraction in LaCOS, for LyC emitters and non-emitters specifically, and for galaxies identified as robust mergers, the broad merger interaction timescales as measured by the merger stage. Figure \ref{fig:rgbs} shows color-composite images of the galaxies identified as mergers and with low merger vote fraction on the same physical scale, with \fesc\ and merger information (merger vote fraction, merger stage) displayed next to each cutout.

\subsection{Merger fractions}

Following visual identification, 48\% of the LaCOS galaxies are identified as secure mergers ($P_{merg}\geq79\%$, corresponding to $3\sigma$), while the rest have low merger vote fraction ($P_{merg}<79\%$). When looking at LyC-emitters specifically, 41\% of the LCEs are identified as secure mergers, while 55\% of the nLCEs are identified as secure mergers.  These values are to be compared with the merger fractions for galaxies in the local Universe \citep[$<5\%$][]{Allam2004,Lotz2008,Conselice2009,Darg2010}, for star-forming galaxies \citep[$\sim10\%$][]{Lotz2008,Kaviraj2015,Duncan2019}, and for highly star-forming galaxies or starbursts \citep[20-50\%][]{Robaina2009,Pearson2019b}. Both LyC emitters and non-emitters show a substantially larger merger fraction than local star-forming galaxies, similar to the fractions found in starburst galaxies. This is likely due to the criteria used to select LyC-emitters in LzCLS, based on either blue UV $\beta$ slopes, large [O$_{\rm{III}}$]/[O$_{\rm{II}}$] line ratios tracing high ionization states, or large star formation rate surface densities $\Sigma$SFR \citep{Flury2022a}, leading to the selection of highly star-forming galaxies. Given mergers can efficiently trigger starbursts, this could readily explain the high fraction of mergers found in both LyC and non-LyC-emitting galaxies in LaCOS.

\subsection{Visually identified merger stages}

\begin{figure}[t]
    \centering
    \includegraphics[width=0.93\linewidth]{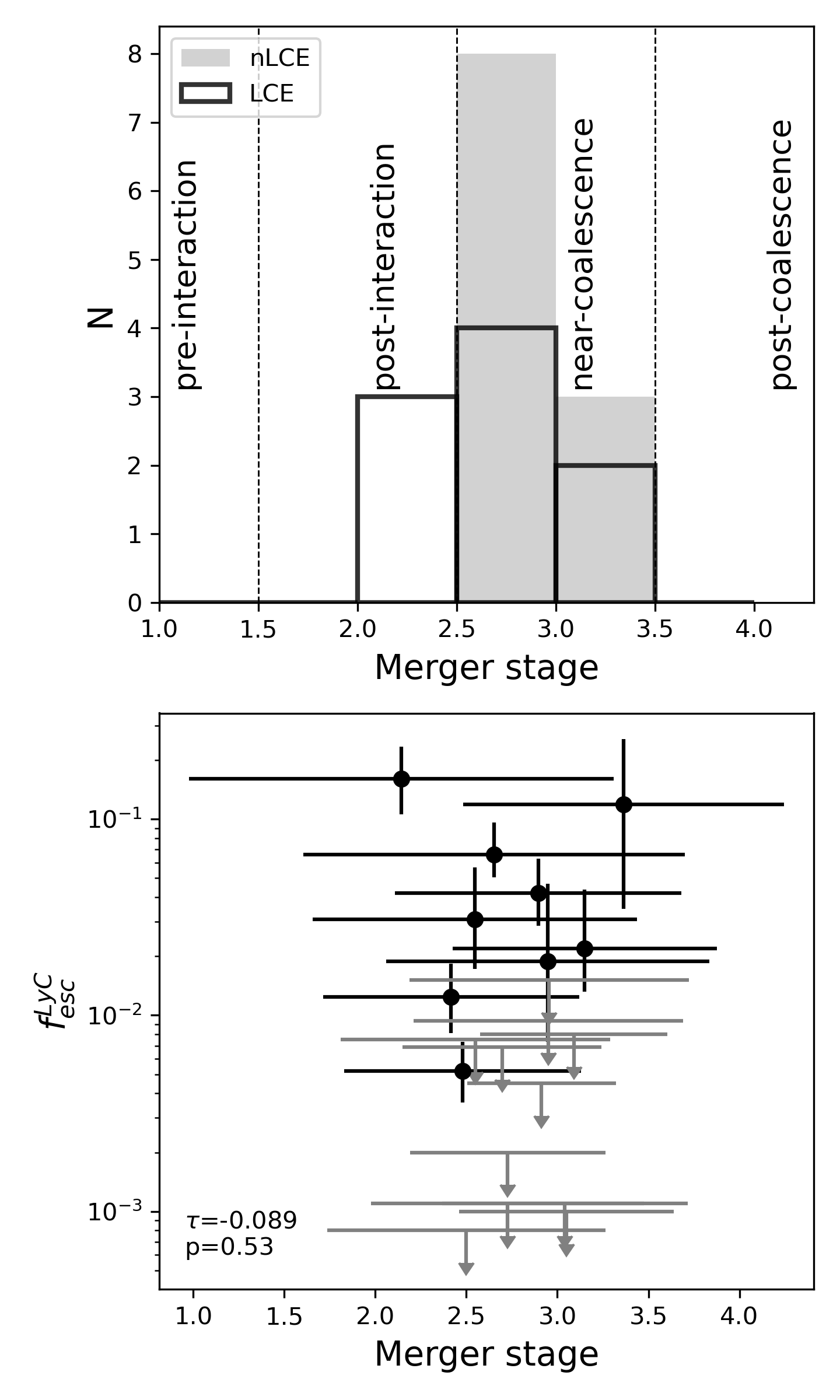}
    \caption{Merger stage and \fesc. Top: Histogram showing the merger stage for LyC-emitting (in black) and non LyC-emitting (in gray) galaxies with $P_{merg}\geq79\%$. The galaxies identified as robust mergers in LaCOS all are at relatively advanced stages of interaction, with most being close to coalescence. Bottom: \fesc\ as a function of the merger stage for galaxies with $P_{merg}\geq79\%$. The Kendall $\tau$ and associated $p$-value characterizing the strength of the correlation are shown on the bottom left corner.  }
    \label{fig:merg_hist}
\end{figure}

Next, we examine the broad timescales associated with the robust merger population identified in LaCOS, as measured by the merger stage introduced in \ref{sec:methods-merg}. Among the 20 mergers in LaCOS, we identify 4 galaxies with morphologies consistent with post-interaction ($1.5\leq s_{merg}<2.5$), and 16 galaxies near-coalescence ($2.5\leq s_{merg}<3.5$). We do not find any pairs in pre-interaction stages or post-coalescence galaxies. Thus, all LaCOS galaxies securely identified as mergers are at advanced stages of interaction, having undergone at least one, or currently undergoing peri-center passages. We visually estimate the separation between nuclei in merging galaxies, and find a median projected separation of $1.4\,$kpc. Comparison to simulated galaxy mergers with similar projected distance in \citet{Patton2024} indicate a relatively short timescale to the merger at such projected distances, with the galaxies likely to merge within a few 100 Myr. In Figure \ref{fig:merg_hist} we show histograms of the merger stage for LCE and nLCEs respectively in the robust merger population. All the non-LyC leaking galaxies identified as mergers are in near-coalescence stages, while LCEs span a slightly larger interaction stage. Nevertheless, we cannot differentiate the samples through a K-S test ($p=0.3$) so that the apparent difference is likely due to small number statistics. We do not find any correlation between \fesc\ and the merger stage, as shown in Figure \ref{fig:merg_hist} (with Kendall $\tau=-0.089$,$p=0.53$). This is likely due to the very narrow range in merger stages sampled by LaCOS, that yields too small of a dynamic range to accurately evaluate the impact of the merger stage on \fesc.

Several studies of the impact of mergers on galaxy properties have found clear trends between enhancements in certain properties and merger timescales \citep{Pearson2025,Ellison2025,Faria2025}. Particularly, in massive galaxies ($M_{*}\geq10^{10}\,M_{\odot}$), merger-driven star formation rate enhancement peaks very close to coalescence, and remains elevated up to 1 Gyr post-merger \citep{Ferreira2025}. Intrinsic LyC production in galaxies is linked to elevated specific SFR ; one may thus expect a similar increase as a function of merger timescale. However, LyC escape also depends on the \hi\ content and covering fraction along the line of sight to UV sources \citep[e.g.][]{Saldana-Lopez2022} which depend on both feedback and tidal gas flows \citep{LeReste2024,Komarova2025}. For this reason, the resulting effect of increased SFR and merger timescale on \fesc\ is hard to extrapolate.
The narrow merger stage distribution observed in LaCOS galaxies, close to coalescence, may indicate that LyC emission in mergers happens in a relatively narrow timescale in a merger interaction \citep[typically a few 100 Myr, see e.g.][]{Patton2024}, at the very end of it.
Alternatively, since the sample selection criteria for LzLCS and LaCOS favor starbursts, and since these are triggered at specific stages of a merger \citep{Faria2025}, they may also preferentially select mergers at certain stages. While we can assert that advanced mergers do facilitate LyC escape in the local Universe, we cannot currently determine that early-stage mergers do not lead to increased \fesc. Surveys investigating the ionizing properties of general galaxy merger samples will be required to draw definitive conclusions on the timescales and merger configurations conducive to LyC emission. 

\subsection{Do robust mergers have specific LyC properties?}

\begin{figure*}[t]
    \centering
    \includegraphics[width=\linewidth]{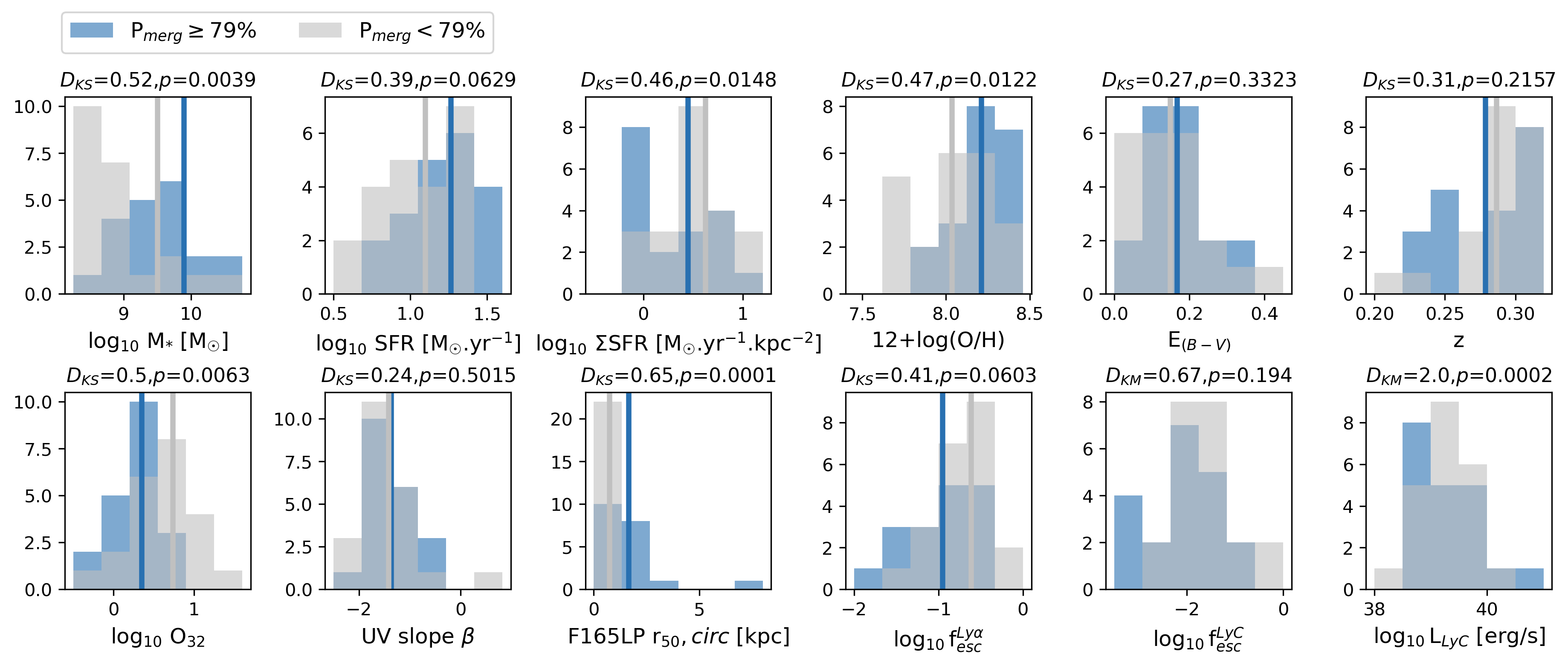}
        \caption{Histograms of various galaxy properties for LaCOS galaxies with $P_{merg}\geq79\%$ (blue) and $P_{merg}<79\%$ (gray). Vertical lines show the mean for each sample. The K-S test (or Log Rank test with Kaplan Meier survival curves, for LyC properties with upper limits) distance statistic and associated $p-value$ for the two samples are shown above each histogram. The robustly identified mergers have properties generally consistent with those of galaxies with low merger vote fractions ($p>0.05$), with the notable exception of $r_{50}$. Robust galaxy mergers have significantly larger radii than their low merger vote fraction counterparts.}
    \label{fig:merg_histprop}
\end{figure*}

Finally, we examine the properties of secure mergers and galaxies with low merger vote fraction, to assess if robust mergers in LaCOS have particular properties. Figure \ref{fig:merg_histprop} displays histograms of the merger and low merger vote fraction galaxies for various galaxy properties, including the stellar mass, star formation rate and star formation rate surface density, oxygen abundance, dust extinction, redshift, O32, UV $\beta$ slope, F165LP $r_{50,circ}$, \fesclya, \fesc\ and LyC luminosity. The values are obtained from \cite{Flury2022a} for most properties, and \cite{LeReste2025} for $r_{50,circ}$ and $\Sigma SFR$ (with star formation rate measured using $H\beta$ fluxes). We note the F165LP $r_{50,circ}$ values are different from those obtained with \texttt{statmorph} presented above as they employ different segmentation maps using integration in circular radii and a custom script. We run K-S tests on all the property distributions, to assess whether or not the robust merger and low-merger vote fraction samples are statistically different. 
For properties with upper limits (\fesc, and $L_{LyC}$), we instead run Mantel log rank tests with Kaplan Meier survival function \citep{Kaplan-Meier1958,Mantel1966}, using the code developed in \cite{Flury2024}. We find that the samples differ in a statistically significant ($p<1.35\times 10^{-3}$) manner only in terms of their UV sizes $r_{50,circ}$ ($p=1\times10^{-4}$) and their LyC luminosity ($p=2\times10^{-4}$). Robust mergers and non-mergers differ tentatively ($1.35e-3<p<0.05$) for $\Sigma SFR$, the $O_{32}$ line ratio, the stellar mass, and metallicity as traced by the oxygen abundance. Specifically, robust galaxy mergers tend to have larger $r_{50,circ}$ (and hence, marginally lower $\Sigma SFR$), larger $L_{LyC}$, marginally lower $O_{32}$ line ratios, marginally larger stellar masses and marginally larger metallicities. 

Examining the \fesc\ properties of robust galaxy mergers, they have escape fractions spanning the range \fesc$=0-16\%$. Instead, the galaxies with low galaxy merger vote fraction span \fesc$=0-49\%$. The galaxies with low merger vote fractions contain the galaxies with the three highest \fesc\ of the LaCOS sample. However, we do not find evidence that their \fesc\ distribution differs significantly from that of the sample of galaxy mergers when running a K-S test ($p=0.09$), meaning the difference likely arises due to the rarity of strong LyC emitters and the small samples sizes. In the following, we discuss the differences between the merger and non-merger samples, and what they could indicate regarding the classification of LyC-emitters.

\begin{figure}
    \centering
    \includegraphics[width=0.9\linewidth]{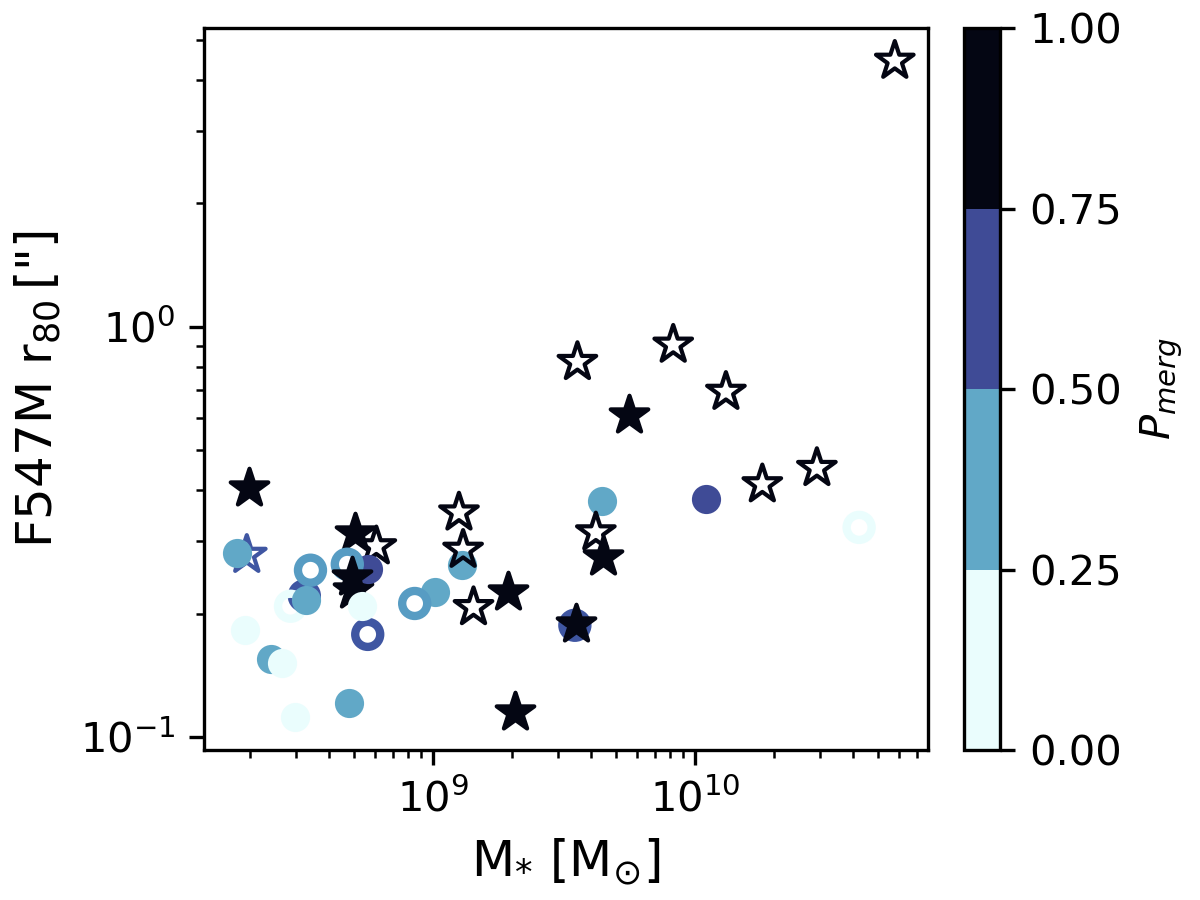}
        \caption{Radius containing 80\% of the light in the F547M filter (similar to rest-frame B-band) as a function of the stellar mass, color-coded by the merger vote fraction. Star marker show galaxies with $P_{merg}\geq79\%$, circle markers show galaxies with $P_{merg}<79\%$. Marker are filled if the galaxy is a LyC-emitter, they are empty otherwise. Compact and low-mass galaxies tend to be voted as mergers less often than massive, extended galaxies in LaCOS.}
    \label{fig:mergfrac_Mstar_r80}
\end{figure}

\section{Discussion} \label{sec:discussion}

\subsection{Are galaxy mergers accurately identified in LaCOS?}
In section \ref{sec:results-mergers} we have shown that all the mergers identified visually with high confidence ($P_{merg}\geq79\%$) in LaCOS are at advanced stages of interaction, having undergone at least one peri-center passage or being close to merging. Additionally, robust mergers and low merger vote fraction samples differ significantly only by their UV sizes as measured by $r_{50}$, and their LyC luminosity, with robust mergers having generally larger sizes and LyC luminosities. Here, we explore potential reasons for these differences and implications for the nature of LyC-emitters. 

On Figure \ref{fig:mergfrac_Mstar_r80} we show the position of mergers and galaxies with low  merger vote fraction in the mass-size diagram, with markers color-coded by the fraction of votes going to merger categories. One may notice that the bulk of galaxies with low merger vote fractions tend to reside in the low-mass ($M_{*}<10^9 M_{\odot}$), small angular size ($r_{80}<0.3"$) corner of the diagram. Interestingly, when examining the merger stages of low merger vote fraction galaxies, we find that when they are voted as mergers, they are consistently identified as post-coalescence galaxies (see Figure \ref{fig:vote}). 

The classification of mergers is generally challenging, and citizen science experiments have shown that even clear galaxy mergers tend to have low merger vote fractions \citep{Lintott2008,Darg2010}. Most galaxies identified in LaCOS may be at post-interaction and pre-coalescence stages, simply because those are the easiest stages to identify visually with certainty, since they present both separate nuclei and tidal features. Therefore, we may expect some of the more compact galaxies in LaCOS to also be mergers, and the merger fractions in LaCOS to be larger than the values derived with our identification scheme. Yet, it is extremely difficult to ascertain which of the galaxies with low merger vote fraction are or are not mergers solely with the HST imaging presented here. Deeper observations could potentially reveal fainter isophotes, and dedicated morphometrics or CNN calibrations adapted to the galaxy property parameter space and specificity of the survey could help refine the identification process. However, and given the data currently at hand, the LaCOS visual merger fractions provide the best lower limits achievable at this stage.

Therefore, two possible options emerge regarding the nature of the objects with low merger vote fraction. The first is that those objects are low-mass, compact star-forming galaxies where stellar feedback is sufficient to allow LyC escape. Simulations have shown that stellar feedback alone can significantly perturb the neutral gas medium in low-mass galaxies, due to their shallow gravitational potential wells \citep[e.g.][]{Rey2022,Trebitsch2017}. The interplay between star-formation and feedback in dwarf galaxies has been proposed to drive episodic increase in star-formation, in a so-called "breathing mode" \citep{Stinson2007,Muratov2015}. In this paradigm, low-mass galaxies go through a starburst cycle that starts with an initial star-forming episode. The newly formed stars lead to both stellar, and eventually, supernova feedback, that expel gas efficiently due to the shallow potential wells, and temporarily stop star formation. As the gas is re-accreted, star formation starts anew and the cycle continues. \cite{Cenci2024} have shown that star formation bursts in simulated galaxies are driven by central gas compaction events, and that the mechanisms responsible for those can be split in two categories. On the one side, merger interactions drive gas compaction and starburst events for most high-mass galaxies ($\rm{M}_{*}>10^{10}\, \rm{M}_{\odot}$), and are also responsible for the most intense and long-lived starbursts at low-mass ($ \rm{M}_{*}\sim10^{8}-10^{10}\, \rm{M}_{\odot}$) \citep{Cenci2024}. On the other side, a significant fraction of starbursting low-mass galaxies (40-50\%) are not undergoing interactions, and instead have bursts occurring due to their star-formation breathing mode \citep{Cenci2024}. Therefore, galaxy mergers may play a major role in facilitating LyC escape mainly from larger-mass galaxies, where feedback alone is not sufficient to clear the neutral interstellar medium. We note similar conclusions have been reached for the role of mergers in \lya-emission, which is also sensitive to neutral gas and dust in the interstellar medium of galaxies \citep{LeReste2025hi,Ren2025}. 
Additionally, this dichotomy of mechanisms, and in particular the breathing-mode of star formation in low-mass galaxies is in agreement with the \cite{Flury2022b} interpretation of LyC measurements as a duty cycle and the \cite{Flury2025} finding that bursty star formation is associated with LyC escape.

Another interpretation may be that some of the compact low-mass galaxies with low merger vote fractions in LaCOS are in fact mergers at coalescence, where the SFR enhancement is the highest \citep{Ferreira2025}. Even in low-mass galaxy samples, interactions account for at least 50-60\% of starburst events, and are responsible for the longest-lived bursts \citep{Cenci2024}. However, low-mass merging galaxies are harder to identify than their higher-mass counterparts, and discerning low-mass mergers at coalescence may be even more difficult. The progenitors of $M_{*}\sim10^{8}\,M_{\odot}$ coalescing systems would have even lower masses $M_{*}<10^{7}\,M_{\odot}$, so that such an interaction would result in extremely faint stellar tidal features. However, since low-mass galaxies are gas-dominated, a potential avenue would be to observe their neutral gas distributions. While the breathing mode of star-formation is expected to yield quasi-isotropic gas outflows \citep{Cenci2024}, mergers result in highly anisotropic and asymmetric gas distributions, that can be probed using the asymmetry applied to 21cm maps \citep{Holwerda2011,Holwerda2025}. 

Such observations have only been carried out in one confirmed LyC-emitting galaxy so far, Haro 11, and showed a spectacularly asymmetric gas distribution \citep{LeReste2024}. However, Haro 11 is one of the LyC-emitters with the largest stellar masses ($\sim10^{10}\, M_{\odot}$) and general conclusions cannot be established from a single observation. Nonetheless, 21cm \hi\ interferometric observations of galaxies analogous to high-z populations, such as green peas and \lya-emitters, have shown that galaxy interactions may be common in those populations \citep{Purkayastha2022,Purkayastha2024,Dutta2024,LeReste2025hi}. The results from 21cm surveys contrast with those from studies that have investigated the large-scale environments of small samples of green pea galaxies \citep{Brunker2022,Kimsey-Miller2024}, finding a large fraction of these objects reside in low-density environments. Additionally, a close-pair search using MUSE observations has found a low fraction of companion galaxies ($\sim10\%$) for green peas \citep{Laufman2022}. Nevertheless, these results do not exclude galaxies in this population being the end-result of galaxy mergers, so that multi-wavelength studies including \hi\ maps could yield a more complete understanding of these objects.

While 21cm interferometric observations have so-far remained mostly limited to the very local Universe ($z<0.05$) where LyC cannot be observed easily due to the low throughput of observation, the SKA is set to observe galaxies in 21cm up to $z\sim1$ \citep{SKAHI}. These observations will make the accurate detection of \hi\ in low-mass mergers feasible at higher redshifts than possible before in the near-future, and thus help the characterization of low-mass, compact, extremely star-forming galaxies. Additionally, the Euclid space telescope is yielding high-resolution imaging ($\sim0.1"$) for a large sky area ($\sim15,000$deg$^2$) \citep{EC:Mellier2025} enabling the detection of large samples of galaxy mergers \citep{EC:LaMarca2025}. Additional investigations into the UV and line emission properties of these large merger samples will help further characterize the ionizing properties of galaxy mergers. Finally, the Habitable Worlds Observatory (HWO), currently in early development phase, may observe LyC emission down to $z\lesssim 0.1$, which would further shed light on the properties of LyC-emitters \citep{Citro2025,Xu2025,McCandliss2025,Carr2025b}. Through coupling with other observations to estimate merger properties, the HWO could provide unique insights into the role of mergers on LyC emission.

\subsection{Implications for merger-driven LyC emission at reionization}
While we have shown galaxy mergers account for $\geq41\%$ of LyC-emitters in LaCOS, and can thus lead to LyC production and escape in the low-z Universe, ultimately we want to determine whether this mechanism could apply at high-z, and specifically at the Epoch of Reionization. 
As already mentioned, mergers are notoriously difficult to detect at high-z, but James Webb Space Telescope observations have recently opened the way to the characterization of mergers in the early Universe. An initial study using morphometrics criteria developed for high-mass, low-z galaxies found that the merger fraction did not evolve with redshift, and merging galaxies did not have properties significantly different from non-mergers \citep{Dalmasso2024}. Additionally, another study using \fesc\ calibrations derived from $z\sim0$ samples and merger identification through morphometrics criteria developed at $z\sim1$ applied to rest-frame UV JWST imaging concluded mergers represent a small fraction of galaxies with high \fesc\ predictions at $z=5-7$ \citep{Mascia2025}.
However, simulations have shown that the use of morphometrics to identify mergers is limited at $z>2$ \citep{Abruzzo2018} and as illustrated in Appendix \ref{app:morph_merg_comp}, morphometrics merger calibrations are specific to the galaxy population and survey considered, likely yielding low accuracy classifications at the Epoch of Reionization. In fact, the results in \cite{Mascia2025} are in agreement with ours, and show that predicted LyC-emitters tend to have low asymmetry in the UV, which is also found in LaCOS. Therefore, the type of objects likely to emit LyC emission, i.e. galaxies with compact, unobscured UV regions and in the case of mergers, systems close to coalescence, might be similar at high redshift. 

Recently, new approaches based on probabilistic photometric pair detection have shed insights into the evolution of merger properties up to $z=11$ \citep{Duan2024,Duan2025}. Those studies have found that the merger fraction increases with redshift up to $z=8$, and that the merger rate increases up to $z\sim6$, after which it flattens. Specifically, a typical galaxy at the Epoch of Reionization will experience on average 6 merger events per Gyr, but merger timescales, while difficult to establish, are also thought to be shorter at high-redshift ($\sim 100$Myr at $z\sim6$) \citep{Duan2025}. The star formation rate of pair galaxies is similar to that of isolated objects, except for very small separations ($<20\,$kpc), where pairs show a small enhancement in SFR \citep{Duan2024}. Thus, enhancements in \fesc\ linked to heightened star-forming activity during interactions may happen on a shorter timescale at reionization than they do in the low-z Universe. However, galaxy pairs show an AGN excess as compared to non interacting galaxies \citep{Duan2024}, which could also potentially play a role in reionization, as AGNs could contribute a fraction of the ionizing photons at $z>6$ \citep{Dayal2020,Trebitsch2023}. Definitive conclusions on LyC emission from mergers at the Epoch of Reionization will require additional work on the LyC properties of galaxy mergers, and their evolution with redshift. In particular, here we have measured the merger fraction among LyC candidates in LaCOS, but the specific criteria used to select galaxies for LyC observations limit range of merger properties probed. 
To fully assess the role of mergers in LyC emission, surveys targeting the ionizing properties of large samples of merging galaxies - particularly those extending to low stellar masses and including AGNs - will be necessary. Combining measurements from such surveys with merger statistics at $z>5$, would allow the estimation of the total ionizing photon contribution from galaxy mergers during the Epoch of Reionization.

\section{Conclusion} \label{sec:conclusion}
We have characterized the morphology and merger properties of LyC-emitting galaxies and candidates at $z\sim0.3$ using rest-frame optical and UV HST imaging from the LaCOS survey \citep{LeReste2025}, which imaged 42 galaxies observed in LyC from the LzLCS survey \citep{Flury2022a}. We find that:
\begin{enumerate}
    \item Lyman Continuum emitting galaxies tend to be compact, both the in rest-frame UV and optical bands. While the criteria used to select galaxies for LyC observations, in particular large $\Sigma SFR$ might contribute to this trend, we find robust anti-correlations between \fesc\ and radii ($r_{20}$, $r_{50}$ and $r_{80}$) in almost all bands sampled by LaCOS photometry. We also find strong anti-correlations between the asymmetry, the clumpiness and \fesc\ in the bluest LaCOS filter available (F150LP). We derive fits between robustly anti-correlated morphometrics and \fesc, with a characteristic rms scatter of $\sim0.5$ dex. We interpret these results as due to LyC photons escaping more easily in compact star-forming galaxies with a small number of centrally-concentrated UV clusters, in agreement with results from other studies using LaCOS data \citep{LeReste2025,Saldana-Lopez2025}. \\  
    \item Both LyC-emitting galaxy samples and samples selected as LyC candidates through properties associated with high \fesc\ (high $O_{32}$, $\Sigma SFR$ or small UV $\beta$ slopes) at $z\sim0.3$ have high merger fractions. Specifically, we find that $\geq48\%$ of LaCOS galaxies and $\geq41\%$ of LaCOS LyC-emitters are galaxy mergers. We do not find statistically significant differences between the \fesc\ properties of robustly identified mergers and galaxies with low merger vote fractions ($P_{merg}<79\%$, equivalent to a $<3\sigma$ confidence merger classification). \\
    \item The galaxy mergers confidently identified in LaCOS are all at advanced stages of interaction (i.e., visually classified as post-interaction or near coalescence), having undergone at least one peri-center passage, and with a median projected separation of $1.4\,$kpc. LyC-emitting and non-emitting mergers have similar merger stage distributions. The range in merger stages is too small to robustly evaluate the impact of merger advancement on \fesc, but our results could indicate that LyC emission during merger interactions happens within a narrow time window ($\lesssim$ few $100\,$Myr) at the end of the interaction. \\
\end{enumerate}

Here, we have presented the most robust estimates to date of the merger fractions in a sample of LyC-emitting galaxies, and have characterized the properties of the identified mergers. While these results offer valuable insight into the connection between galaxy interactions and LyC escape, the classification of mergers remains intrinsically challenging, and the criteria used to select our sample limit inferences on the role mergers play in facilitating LyC emission. Moving forward, dedicated multi-wavelength surveys targeting mergers across a broad dynamical range will be crucial to improve estimates of the merger timescales and configurations during which LyC leakage is most likely to occur. Notably, 21cm \hi\ emission mapping of galaxies selected through \fesc\ calibrations at lower redshifts would significantly enhance classification accuracy by helping identify post-coalescence, compact mergers likely to be missed in optical and UV imaging. New and upcoming facilities such as Euclid, the SKA, and the Habitable Worlds Observatory will be instrumental in solving the role of mergers on LyC emission. When combined with detailed merger statistics at $z>5$, these data will enable a comprehensive evaluation of the contribution of galaxy mergers to the Epoch of Reionization.

\section{Data availability}
The HST data used for the analysis in this manuscript is publicly released at the Barbara A. Mikulski Archive for Space Telescopes (MAST) as a High Level Science Product, accessible via \dataset[doi:10.17909/j4qd-ev76]{\doi{10.17909/j4qd-ev76}} 
and \url{https://archive.stsci.edu/hlsp/lacos/}. 

\begin{acknowledgments}
This research is based on observations made with the NASA/ESA Hubble Space Telescope obtained from the Space Telescope Science Institute, which is operated by the Association of Universities for Research in Astronomy, Inc., under NASA contract NAS 5–26555. These observations are associated with HST GO programs 17069, 14131 and 11107. This research has made use of the NASA/IPAC Infrared Science Archive, which is funded by the National Aeronautics and Space Administration and operated by the California Institute of Technology. Based on observations collected at the European Southern Observatory under ESO programme ID 179.A-2005 and on data products produced by CALET and the Cambridge Astronomy Survey Unit on behalf of the UltraVISTA consortium. ALR thanks Rui Marques-Chaves, Stephan McCandliss, Laura Pentericci and Axel Runnholm for their contribution to merger identification in LaCOS. ALR and MSO acknowledge support from HST GO-17069. KBM acknowledges partial support for this work from the following grants: NSF IIS 2006894 and NASA Award \#80NSSC20M0057.
\end{acknowledgments}

\software{numpy \citep{numpy}, astropy \citep{astropy1,astropy2,astropy3}, matplotlib \citep{matplotlib}, photutils \citep{photutils}, scipy \citep{scipy}, statmorph \citep{Rodriguez-Gomez2019}, linmix \citep{Kelly2007}, sep \citep{Barbary2016}, histogram \citep{Flury2022a}, KaplanMeier \citep{Flury2024}, petrofit \citep{Geda2022,petrofitZenodo}.}

\facilities{HST,IRSA}
\bibliography{bibliography}{}
\bibliographystyle{aasjournal}

\appendix
\section{Example image panels for classification}
\label{app:visual_class_panel}
\begin{figure}
    \centering
    \includegraphics[width=0.65\linewidth]{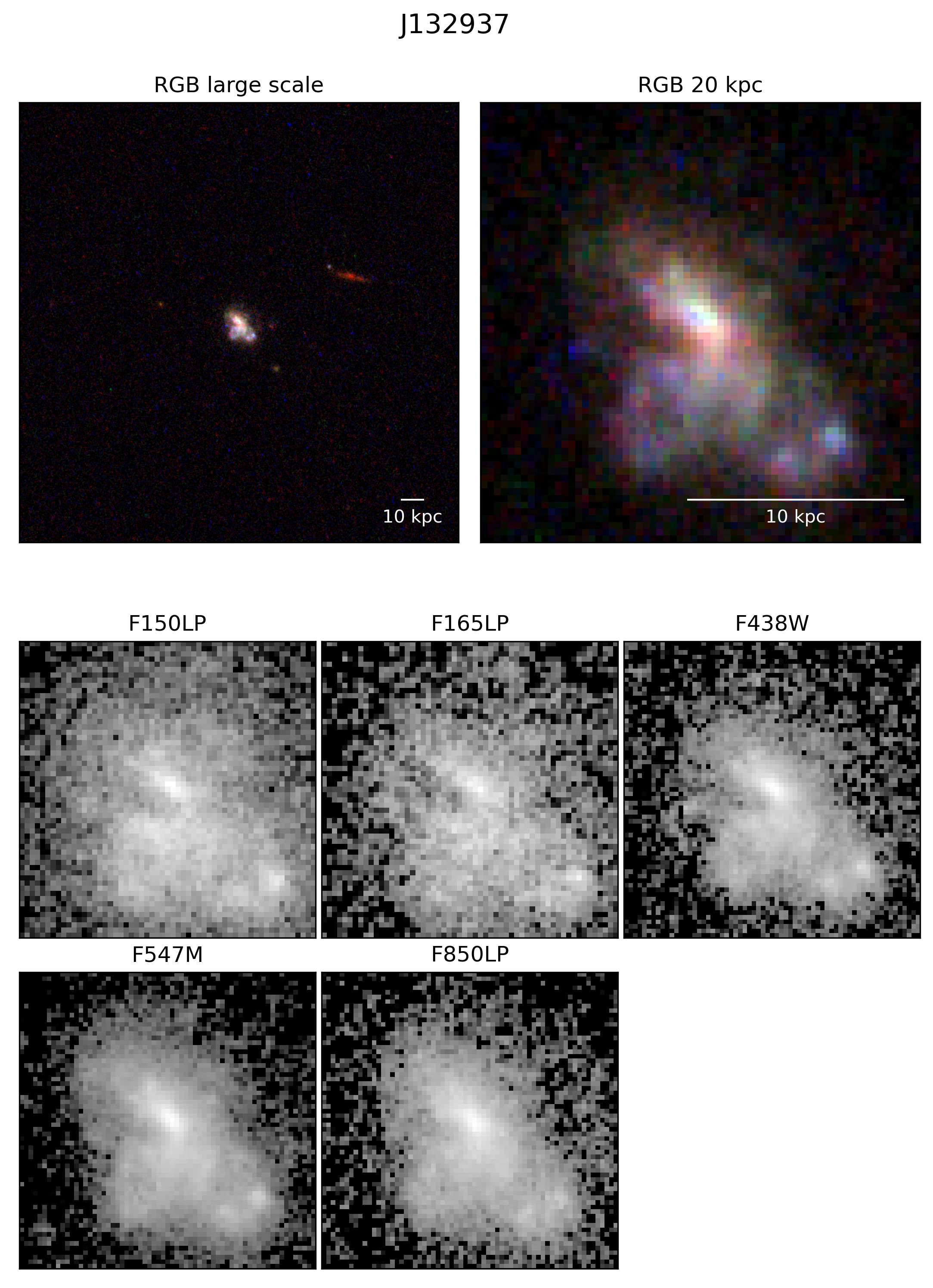}
    \caption{Example thumbnail shown to classifiers for merger identification.}
    \label{fig:mergclass_thumb}
\end{figure}

On Figure \ref{fig:mergclass_thumb} we show the example of an image panel presented to classifiers for the visual classification of galaxy J132937. This galaxy is classified as a robust galaxy merger near-coalescence with $P_{merg}=0.92$ and $s_{merg}=2.7$. The two top panels show optical RGB images with scales chosen to maximize the visibility of structures across all filters. The top left panel shows a 24"-wide image, the top right panels show a 20kpc cutout of that image. The bottom panels show the same 20kpc cutouts for all available filters. 

\section{Comparing visual with non-parametric morphological merger identification}
\label{app:morph_merg_comp}
\begin{figure}
    \centering
    \includegraphics[width=0.9\linewidth]{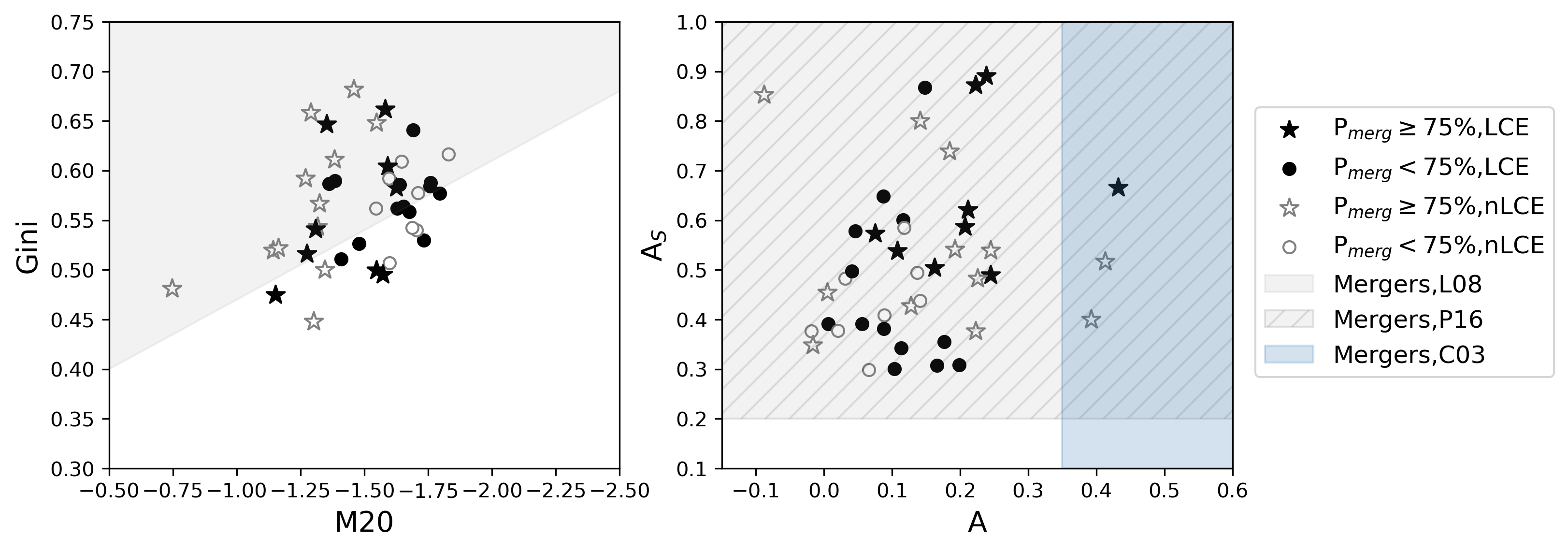}
    \caption{Visual merger classification as compared to morphometrics selection using F547M and F850LP images (closest to rest-frame B/g- and r/I-bands). The left panel shows the position of LaCOS galaxies on the Gini-$M_{20}$ diagram using F547M images, with the parameter space for merger in \cite{Lotz2008} overlaid in gray. The right panel shows LaCOS galaxies on the shape asymmetry and asymmetry diagrams using F850LP images, with the parameter space for merger in \cite{Conselice2003} overlaid in blue and that in \cite{Pawlik2016} overlaid in gray.}
    \label{fig:enter-label}
\end{figure}

Figure \ref{fig:enter-label} presents the position of LaCOS galaxies on morphometric diagrams. Specifically, we show planes defined by merger selection criteria using the Gini and $M_{20}$ coefficient \citep[$G>-0.14*M_{20} + 0.33$][]{Lotz2008}, the asymmetry \citep[A$\geq$0.35][]{Conselice2003} and the shape asymmetry \citep[$A_s\geq0.2$][]{Pawlik2016}. The latter parameter is more sensitive to faint isophotes than the traditional asymmetry, and has been developed to facilitate the identification of post-coalescence mergers. There is a large degree of variation in the galaxies identified or not as mergers via typical morphometrics criteria. While the asymmetry criterion selects only galaxies with high merger vote fraction, it fails to detect most of the robust mergers identified visually. The Gini-$M_{20}$ selection identifies about 75\% of the robust mergers, but also selects 64\% of the low vote fraction  mergers. Finally, the selection based on shape asymmetry selects all of the LaCOS galaxies as mergers. These criteria were developed on B-band \citep{Lotz2008}, I-band \citep{Conselice2003} and r-band \citep{Pawlik2016} images of relatively massive galaxies more representative of the general $z\sim0-1$ galaxy population, likely resulting in the low classification accuracy observed for LaCOS galaxies. This highlights the importance of considering the specifics of a given sample when selecting a merger classification scheme.

\section{Morphometrics and merger parameters}
\label{app:table}
We provide a representative description of the data produced as part of the analysis and available in the online version of this manuscript in Table \ref{tab:morph}. The table contains the merger parameters resulting from the classification scheme described in \ref{sec:methods-merg} and non-parametric morphological parameters calculated using images in all available HST filters in the LaCOS survey (ACS/SBC/F150LP, ACS/SBC/F165LP, WFC3/UVIS/F438W, WFC3/UVIS/F547M, and WFC3/UVIS/F850LP or ACS/WCS/F850LP).

\begin{deluxetable}{cccc}[h]
\label{tab:morph} 
\tablecaption{Morphometrics and merger parameters for LaCOS galaxies.}
\tablehead{
   \colhead{Row number} & \colhead{Units} & \colhead{Label} & \colhead{Description}
}
\startdata
1 & --- & ID & Source identifier \\
2 & deg & RAdeg & Right Ascension in decimal degrees (J2000) \\
3 & deg & DEdeg & Declination in decimal degrees (J2000) \\
4 & --- & z & Redshift\\
5 & --- & fvotemerg& Fraction of votes toward a merger category\\
6 & --- & fvotenmerg& Fraction of votes toward the non-merger category\\
7 & --- & fvotensure& Fraction of votes toward the not-sure category\\
8 & --- & mergstage& Merger stage\\
9 & --- & mergstagedev& Merger stage standard deviation\\
10 & --- & asymmetry(f150lp)& Asymmetry in the F150LP filter\\
11 & --- & concentration(f150lp)& Concentration in the F150LP filter\\
12 & --- & gini(f150lp)& Gini parameter in the F150LP filter\\
13 & --- & m20(f150lp)& M20 parameter in the F150LP filter\\
14 & pix & r20(pix)(f150lp)& Radius including 20\% of the light in the F150LP filter\\
15 & pix & r50(pix)(f150lp)& Radius including 50\% of the light in the F150LP filter\\
16 & pix & r80(pix)(f150lp)& Radius including 80\% of the light in the F150LP filter\\
17 & --- & smoothness(f150lp)&Clumpiness parameter in the F150LP filter\\
18 & --- & shapeasymmetry(f150lp)&Shape Asymmetry in the F150LP filter \\
...\\
\enddata
\tablecomments{Table \ref{tab:morph} is published in its entirety in the electronic 
edition of the {\it Astrophysical Journal}.  A portion is shown here 
for guidance regarding its form and content.}
\end{deluxetable}

\end{document}